\documentclass[letter,12pt]{article}

\usepackage{graphicx}

\usepackage[style=apa]{biblatex}
\usepackage[ruled,vlined]{algorithm2e}

\usepackage{amsmath}
\usepackage[left=3cm, right=3cm, top=3cm]{geometry}

\usepackage{url}

\newtheorem{hypothesis}{Hypothesis}

\begin{document}

\title{\vspace{0cm} Forecasting skill of a crowd-prediction platform: A comparison of exchange rate forecasts
}

\author{Niklas V. Lehmann\thanks{Technical University Bergakademie Freiberg, Researcher at the Chair for General Economics and Macroeconomics. Contact: Niklas-Valentin.Lehmann@vwl.tu-freiberg.de \\
Schlossplatz 1, 09599 Freiberg, Fakultät 6, GERMANY}
        }

\date{\textit{February 2025}}
\maketitle

\begin{abstract}
Open-online crowd-prediction platforms are increasingly used to forecast trends and complex events. In this analysis, exchange rate predictions made on Metaculus are compared to predictions made by the random-walk, a statistical model considered very hard-to-beat. The crowd-prediction proves to be less accurate than the random-walk. By using the random-walk as a benchmark, this analysis provides a rare comparison of online crowd-prediction platforms with traditional forecasting techniques.

\noindent
\end{abstract}
\vspace{0.5cm} \noindent

\noindent \textbf{Keywords}:  Crowd-prediction, Wisdom of crowds, Random-walk, Forecast accuracy
\vspace{6pt}

\noindent \textbf{JEL}: C53, C58

\vspace{12pt}

\noindent \textbf{Data Availability \& Conflicts of interest:}
This paper contains all relevant information to replicate and build upon the results. Relevant data on the Metaculus prediction can be obtained via the API that Metaculus provides. I did not receive any funds or other benefits for the research presented and therefore have no conflicts of interest. None of the views presented here are attributable to Metaculus.

\vspace{12pt}

\noindent \textbf{Acknowledgements:}
I am indebted to Robert L. Czudaj and Nikos Bosse for their valuable comments on an earlier draft of this paper.

\thispagestyle{empty}
\clearpage
\addtocounter{page}{-1}

\setlength{\baselineskip}{18pt}


\section{Introduction}

Forecasts from crowd-prediction platforms - online platforms that allow anyone to predict outcomes of public questions - are increasingly seen as sources of foresight and evidence. Forecasts from open-online crowd-prediction platforms have been featured in European Central Bank reports (\cite{ECBnews}) and several news sites such as The Economist, Forbes, The Washington Post and Vox have started to incorporate crowd-predictions.\footnote{See e.g. (\cite{economist_article})} Moreover, it is widely accepted that crowd-prediction is a valuable tool that can support policy decisions (\cite{tetlock2017bringing}). The US intelligence community, the Virginia Department of Health and Our World in Data have leveraged crowd-predictions to inform their research (\cite{tetlock2014forecasting}, \cite{metaculus2022}, \cite{metaculus2023}).
Crowd-predictions have also been directly used for research purposes, e.g. to compare them to the prediction capabilities of large language models  (\cite{schoenegger2023large}).

However, it is yet uncertain exactly how much confidence we should place in the predictions of crowd prediction platforms. Whilst there is an abundance of evidence that shows that they provide forecasts that are more accurate than random guessing (\cite{petropoulos2022forecasting}, \cite{atanasov2017distilling}),
there is little evidence on the comparative accuracy of crowd-predictions.
Only a limited number of studies explore whether crowd-predictions are more accurate than traditional forecasting methods in relevant areas.
\textcite{farrow2017} demonstrates that crowds produced superior short-term predictions for flu cases compared to statistical models. \textcite{mcandrew2024assessing} shows that crowds outperform statistical models in forecasting monkeypox outbreaks; however, statistical methods are more accurate in 1-week-ahead-forecasts. Meanwhile, \textcite{mcandrew2024chimeric} indicates that combining crowd-predictions with statistical models can enhance the accuracy of epidemic forecasts. Additionally, \textcite{SuperforecastingFed2023} reveals that crowd-predictions provided more accurate interest rate forecasts than the CME FedWatch Tool. 

This study assesses the accuracy of crowd-predictions from the Metaculus platform on questions related to exchange rates, where an objective and well-studied benchmark, the random-walk without drift, exists.
Furthermore, the studied predictions had real-world importance. They were supposed to create decision support for operational needs of humanitarian agencies by identifying potential upcoming conflict zones and economic crises around the world.\footnote{See \texttt{www.metaculus.com/questions/11505/economic-trouble-15-currency-depreciation/}}
The result of this study is that the random-walk without drift provides significantly more accurate predictions than the crowd. 



This paper proceeds as follows: The next section provides a literature review of both crowd-prediction in general as well as forecasting exchange rates. Section 3 describes the historical data and the crowd-prediction platform utilized in this study. Section 4 describes the studies methodology. Section 5 then presents results, which are briefly discussed in section 6. Section 7 concludes.


\section{Literature Review}






\subsection{Crowd-prediction}





There are many ways to elicit individual forecasts and combine them to form a consensus forecast, such as simple averages. Different ways of distilling the wisdom of crowds have been suggested and tested (\cite{atanasov2022crowd}, \cite{armstrong2001combining}). Among the most well studied are prediction markets (\cite{arrow2008promise}, \cite{hanson2003combinatorial}),  forecasting tournaments,  and prediction polls (\cite{atanasov2017distilling}). 
Crowd-prediction, in the form of prediction markets, has been successfully employed at large companies to aid decision-making (\cite{cowgill2015corporate}).


Open online crowd-prediction, hereinafter just 'crowd-prediction', is a type of forecast that results from combining predictions made by multiple forecasters via a shared online platform. Crowd-prediction platforms are similar to forecasting competitions in that they involve multiple participants, and while there are often monetary rewards involved, most forecasters are primarily driven by the desire to establish their reputation and prestige by winning competitions and demonstrating a history of accurate predictions (\cite{mellers2014psychological}).


These crowd-prediction platforms have demonstrated impressive foresight across a wide range of questions in recent times (\cite{tetlock2014forecasting}, \cite{tetlock2017bringing}). 
According to \textcite{nofer2015crowds}, the online stock prediction community has consistently performed better than professional analysts when it comes to forecasting stock returns. \textcite{brown2019sportTipster} find that an online community of amateur tipsters outperformed bookmakers in real-money sports bets, when predictions of tipsters are properly combined. Additionally, \textcite{sjoberg2009all} shows that crowds have been successful in accurately predicting political and geopolitical events. This particular application of crowd-prediction is widely utilized due to the limited alternatives available for forecasting complex events. Moreover, the paper demonstrates that crowd-predictions are as accurate as expert predictions in this domain. \textcite{katz2017crowdsourcing} find that crowds have been the best source of foresight regarding Supreme Court decisions in the United States. 


\subsection{Forecasting exchange rates}

Throughout history, numerous endeavors have been made to forecast exchange rates. These endeavors have revealed the challenges associated with predicting exchange rates (\cite{cornell1978efficiency}, \cite{giddy1975random}). It has been observed that the \textit{random-walk without drift} is not systematically outperformed by any other (more sophisticated) statistical model (\cite{rossi2013exchange}). Predicting exchange rates is a challenging task due to the potential for profit if future exchange rate movements were known in advance. By buying currency at a low price and selling it at a high price, easily predictable movements should largely vanish. Consequently, information about future exchange rates should already be factored into current rates, leaving little room for predictable drift (\cite{giddy1975random}).


 However, the behavior of exchange rates cannot be entirely explained by economic theory (\cite{meese1983empirical}, \cite{rossi2013exchange}). Despite the extensive research on the topic, there is still limited understanding of why exchange rates move in the manner they do and why they may not adhere to concepts such as interest parity (\cite{chinn2004monetary}, \cite{kilian2003so}, \cite{engel2005exchange}). This phenomenon is commonly referred to as the Meese-and-Rogoff puzzle (\cite{meese1983empirical}).
Although there seem to have been modest successes at predicting exchange rates beyond random fluctuation (\cite{li2015predicting}, \cite{beckmann2020exchange}), researchers have struggled to reliably outperform the random-walk with statistical models.  
As a result, the prediction produced by the random-walk serves as a benchmark, representing the best-known approach.

Judgmental forecasting is also used to predict exchange rates, mostly in the form of surveys. These surveys, typically provided by firms such as Consensus Economics or FX4casts, collect forecasts from economists on a quarterly basis for specific questions such as: "What will be the value of the Euro (measured in USD) on January 1st, 2025?" Forecasters participating in these surveys provide a single point prediction for these exchange rate questions.
\textcite{macdonald2009exchange} find evidence suggesting that certain forecasters possess valuable insights into future exchange rates, while others do not. \textcite{onkal2003professional} also observe that, on average, experts outperform amateurs in short-term exchange rate forecasts, whilst \textcite{leitner2006} find the opposite.

\section{This study}

This study assesses the accuracy of crowd-predictions from the Metaculus platform on questions related to exchange rates, where the random-walk provides an objective, difficult-to-beat and well-studied benchmark that can be constructed post-hoc. The central research question is: 

\begin{quote}
\textbf{Is the crowd-prediction as accurate as the random-walk in forecasting exchange rates?}
\end{quote}


\begin{hypothesis}
The crowd-prediction and the random-walk produce the same error in forecasting exchange rates. 
\end{hypothesis}

Accuracy shall be measured via the \textit{squared error}, often also known as the \textit{brier score} (\cite{brier1950verification}). 
The squared error is used primarily because Metaculus forecasters were also assessed with the squared error. Furthermore, the squared error is a strictly proper scoring rule. Strictly proper scoring rules have the property that they are maximized in expectation by the true value. This means that forecasters have an incentive to carefully assess the question and provide their honest answer.\footnote{This assumes linear von Neumann-Morgenstern utility functions (\cite{gneiting2007strictly}) and a host of other conditions that are not necessarily met at Metaculus. However, this is not a problem for our analysis because we just assess errors ex post.} 
The squared error is described in equation \ref{eq:squared_error} and \ref{squared_error}. If the event occurred $k$ is 1, otherwise $k$ is 0. The prediction made shall be $p_t$. 

\begin{align}
     y(k=0) &=  p_t^2 \label{eq:squared_error} \\ 
     y(k=1) &=  (1-p_t)^2 
    \label{squared_error}
\end{align}


\begin{hypothesis}
The crowd-prediction produces a lower error in forecasting exchange rates than random guessing.  
\end{hypothesis}

Furthermore, we ask whether the crowd possesses any foresight at all. Therefore, we compare the crowd with an imaginary random guesser, who always predicts 50\%. Such predictions are useless, but still yield a squared error of $0.25$. 
We are also interested in systematic differences between the crowd-prediction and the random-walks prediction because
such differences may reflect information that is contained in one forecast but not in the other. This is reasonable as the human forecasters have access to news and other sources.
If so, we may be able to retrieve a more accurate forecast by combining the predictions of random-walk and the crowd, as 
\textcite{mcandrew2024chimeric} do for epidemic forecasts.


\begin{hypothesis}
The crowd-prediction is not systematically different from the random-walk prediction.  
\end{hypothesis}

\section{Data}

\subsection{Metaculus forecasting platform}


Metaculus is a crowd-prediction platform working to improve human reasoning and coordination on topics of global importance. As a Public Benefit Corporation, Metaculus largely provides forecasts publicly. Metaculus features questions on a wide range of topics. As of 2025,  over ten thousand questions have been submitted, over half of which have been evaluated since the platforms inception in 2015. Over 2 million individual forecasts have been made on the platform by thousands of active users. 
This method for crowd-prediction is a forecasting tournament (\cite{tetlock2014forecasting}) and not to be confused with other ways of eliciting crowd opinion such as prediction markets.  
Metaculus publishes a combined forecast that uses the track record of forecasters to weight predictions, giving more weight to historically accurate forecasters. This forecast is called the \textit{Metaculus prediction} and serves as the 'crowd-prediction' in this study. \footnote{Metaculus chooses to disclose how exactly their aggregation mechanism works. The Metaculus prediction is usually hidden for as long as questions are still open for predictions. The community prediction \textit{can} also be hidden for some period after the opening of a new question. This feature may limit early groupthink, yet disadvantages predictors which predict on questions early.}

\begin{figure}
    \centering
    \includegraphics[width=.85\textwidth]{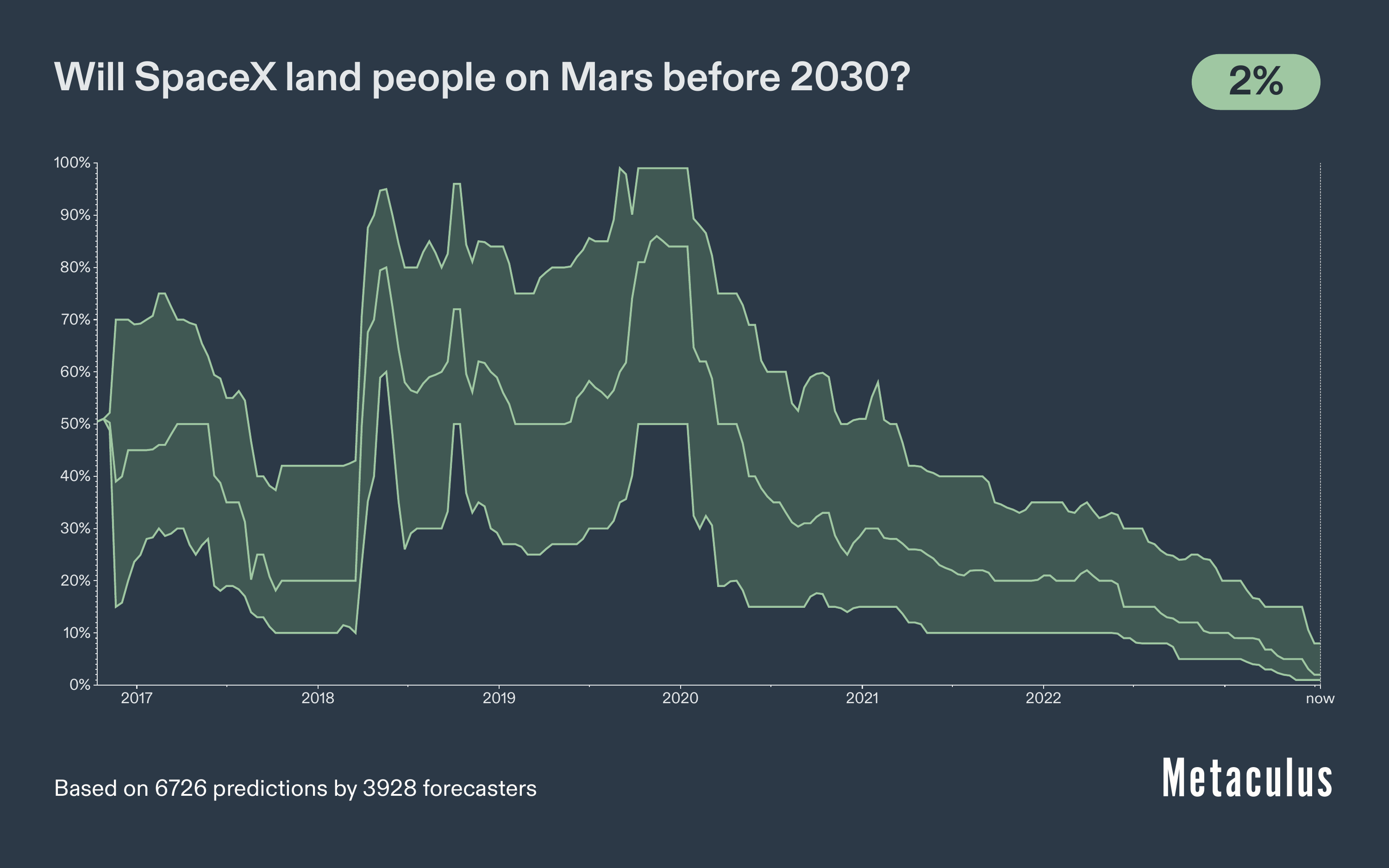}
    \caption{Metaculus interface to explore how predictions on a question have evolved}
    \label{fig:Metaculus_prediction}
\end{figure}
\begin{figure}
    \centering
    \includegraphics[width=.85\textwidth]{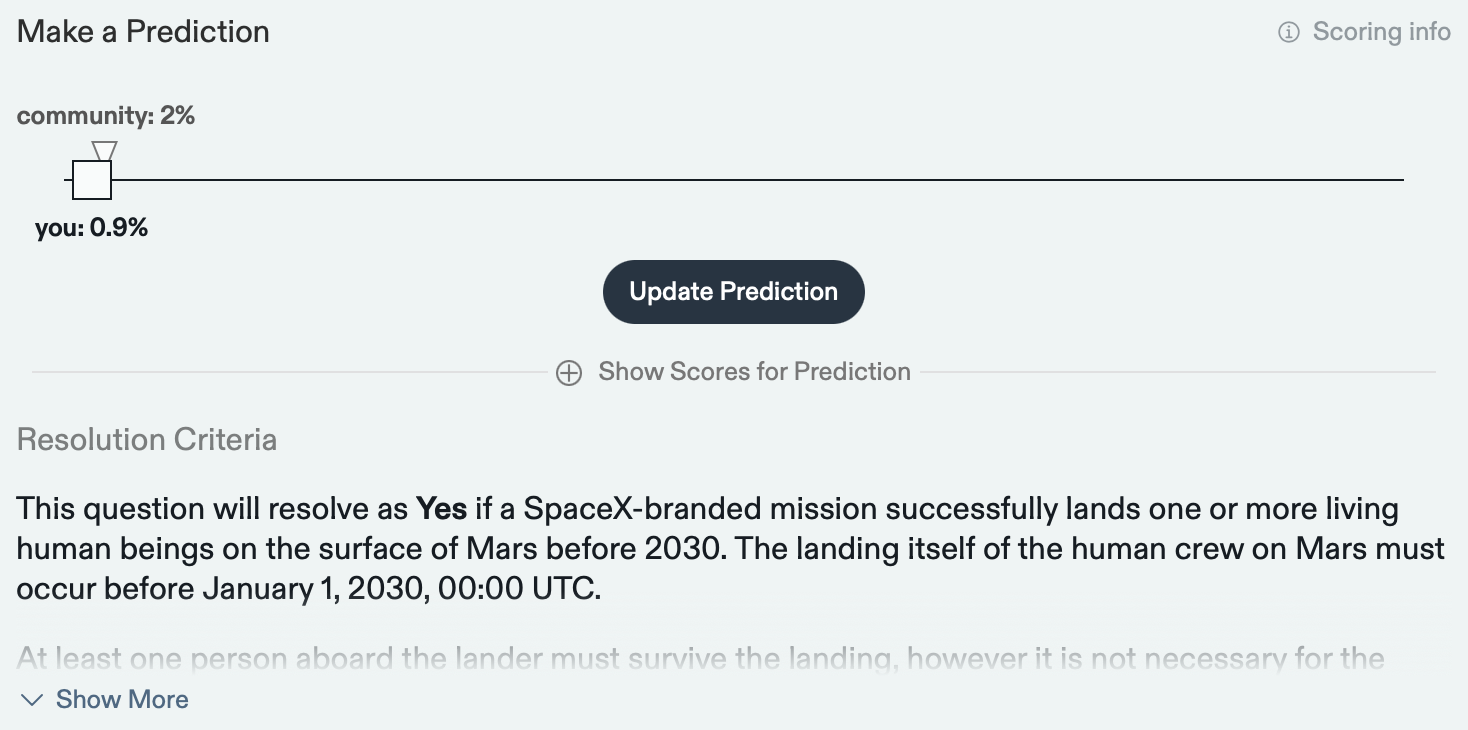}
    \caption{Metaculus prediction interface for the SpaceX-question from figure \ref{fig:Metaculus_prediction}}
    \label{fig:Metaculus_prediction_interface}
\end{figure}

Forecasters submit predictions at their leisure, and can make as many predictions at any point in time as they like. Forecasters use sliders to report their prediction. The prediction interface is depicted in figure \ref{fig:Metaculus_prediction_interface}. Forecasters also have access to the median prediction as well as quartiles. Figure \ref{fig:Metaculus_prediction} shows how forecasters see other forecasters predictions.

\subsection{Forecasting question series on exchange rates}

I compiled all exchange rate questions on Metaculus that can be directly compared to a prediction made by the random-walk. These include 12 questions from a  question series that ran between June and December 2022, as well as 2 questions regarding a potential British Pound parity with the dollar. Table \ref{tab:questions} provides a comprehensive list of the relevant exchange rate questions that were featured on Metaculus.

\begin{table}[]
    \centering
    \begin{tabular}{c|ll}
        1 &  Euro to depreciate by $>15\%$ in 2022? & $k=0$\\ 
        2 &  Indonesian Rupiah to depreciate by $>15\%$ in 2022? & $k=0$\\
        3 &  Thai Baht to depreciate by $>15\%$ in 2022? & $k=0$\\
        4 &  Russian Ruble to depreciate by $>15\%$ in 2022? & $k=1$\\
        5 &  Turkish Lira to depreciate by $>15\%$ in 2022? & $k=0$\\
        6 &  Polish zloty to depreciate by $>15\%$ in 2022? & $k=0$\\
        7 &  Brazilian Real to depreciate by $>15\%$ in 2022? & $k=0$\\
        8 &  Mexican Peso to depreciate by $>15\%$ in 2022? & $k=0$\\
        9 &  Indian Rupee to depreciate by $>15\%$ in 2022? & $k=0$\\
        10 &  Pakistani Rupee to depreciate by $>15\%$ in 2022? & $k=0$\\
        11 &  Chinese Yuan Renminbi to depreciate by $>15\%$ in 2022? & $k=0$\\
        12 &  British Pound to reach market parity with the dollar by 2017? & $k=0$\\ 
        13 &  British Pound to reach market parity with the dollar by 2023?& $k=0$\\ 
    \end{tabular}
    \caption{Studied questions from the crowd-forecasting platform}
    \label{tab:questions}
\end{table}

 All exchange rates in this study are based on the US dollar as the reference currency. Question 12 opened on 2016-07-09 and question 13 opened on 2022-09-29.
 The questions in this study were classified as resolved 'Yes' ($k=1$) if the depreciation threshold was reached, or 'No' ($k=0$) if the depreciation threshold was not reached by the time the questions closed on December 31st 2022.
 Over a period of six months, a total of 61 amateur forecasters participated in the question series, collectively submitting 2453 individual predictions.\footnote{Since this study is purely observational and not conducted in a lab-setting, no additional information about forecasters is available.} On average, there were 144 total predictions per question.\footnote{The question series can be explored online at \\ \texttt{www.metaculus.com/questions/11505/economic-trouble-15-currency-depreciation/}.} As an incentive for their participation, forecasters received a small monetary compensation based on the accuracy of their predictions. The forecasters were rewarded based on their squared error. A 2500\$ prize pool was awarded to accurate forecasts across a larger set of 64 questions, which included the questions from 1 to 12. The average amount of money awarded per predictor was below 10\$. However, only the most accurate forecasters could claim payments. The predictions on questions 12 and 13 were not monetarily incentivized.

\section{Methodology}


In order to compare the Metaculus prediction, which is discussed in the prior section, to the random-walks predictions, we need to generate the latter. 
Since this study seeks to collect predictions for whether a currency will depreciate to a certain value by a certain time, a Monte Carlo simulation was used to collect these predictions. 
The random-walk, described in equation \ref{randomwalk}, is a model that simply projects the normal distribution with the historic standard deviation into the future. 

\begin{equation}
    x_{t+1} = x_{t} + \epsilon_t, \quad \text{with} \ \epsilon_t \sim \mathcal{N}(0,\sigma_h^2)
    \label{randomwalk}
\end{equation}  

$x_t$ shall be the value of a currency at time $t$, as measured in US-dollars. The next time steps value is defined as the current value plus some random change ($\epsilon$), which is normally distributed with the historic standard deviation of the exchange rate series.

The financial data on exchange rates was gathered from Yahoo Finance using the quantmod-package in \verb|R|.\footnote{The data used to resolve the crowd-prediction questions was sourced from \texttt{xe.com}. In other words, the crowd-prediction is based on other financial data then the random-walk-prediction. The difference between the two is small and negligible in comparison to the exchange rate movements and uncertainty in exchange rate expectations displayed.} 
Exchange rate data was collected on a daily basis from January 5, 2022, until December 31, 2022, which encompassed the resolve time of the questions. There was an exception for the questions specific to the British Pound, which opened earlier. For these cases, exchange rate data was gathered starting from January 1, 2015.
 Algorithm 1 describes how the predictions were generated.\footnote{Exchange rate fluctuations are not observed on weekends - a fact that is brushed over in the Monte-Carlo-Simulation. This is insofar relevant as currencies cannot depreciate below the threshold on weekends. If a currency is close to the depreciation barrier, but there are no more days in which the market is open in the year, the algorithm would still assume a potential depreciation of the currency on each day. However, this situation has not turned up in this analysis.} 

\begin{algorithm}[H]
\caption{The random-walks prediction}
\KwIn{Historical exchange rate time series; Date from which to forecast}
\KwOut{Probability of currency depreciation}

Step 1: The historical variance $\sigma_h^2$ of the exchange rate is calculated based on the available data for that specific day. That is, if the probability of the Euro depreciating below the specified threshold is to be predicted for Oct. 30th 2022, then the historical standard deviation is calculated using the exchange rate values (variance) from January 5th on up to the 29th of October.\;
Step 2: Ten thousand paths of the required length (up until December 31) are generated using a Monte-Carlo simulation. The next days value is determined by equation 1, whereas $\epsilon$ is a random draw from a normal distribution with a variance equal to the historical variance $\sigma_h^2$ computed in the previous step.\;
Step 3: To derive the probability that a currency will depreciate below the specified threshold, the number of simulated paths where this occurs (at some point), is divided by 10000 (the total number of paths).\;

\end{algorithm}

Most importantly, the approach arrives at a probability or forecast that anyone would have been able to easily generate in real time. Only information available at the specific day is used to make forecasts. The forecasts reflect pseudo-out-of-sample-performance of the random-walk. Thereby, the forecasts generated by this method provide a fair benchmark for the crowd-prediction.



In order to evaluate the performance of the crowd-prediction and the random-walk model, and test hypothesis 1, forecasts were compared on a question-by-question basis. 
For each individual question, and corresponding exchange rate time series, the squared error of the crowd-prediction and the random-walk model's predictions were computed. Errors of the crowd-prediction and random-walk were then summed up across questions and divided by the number of questions to arrive at the mean squared error.\footnote{Since some questions resolved early, the number of questions changes across time.} The method with the lower mean squared error is more accurate.
In order to assess whether differences in accuracy between the two methods are merely a chance result, a Diebold-Mariano test is deployed (\cite{diebold2002comparing}). 
The Diebold-Mariano test takes the difference between the errors produced by either forecasting technique and uses a simple z-test to check whether the difference in error can be explained by variation around 0.  
This plain version of the test would assume a stationary time series, i.e. that the differences in error are normally distributed draws with a mean of 0 and an estimated variance that is constant. 
The situation here is different, since errors are clearly autocorrelated. Therefore, this analysis employs heteroskedasticity-and-autocorrelation-robust-standard-error estimates, as suggested by \textcite{diebold2015comparing}.
Furthermore, we test the hypothesis 2, that the Metaculus prediction is no more accurate than a random guesser, by taking the difference between the mean squared error of the crowd and $0.25$ (the imaginary random guessers error) and deploying a Diebold-Mariano test. If this test turns out to be negative, we can conclude that the crowd does possess foresight regarding exchange rate movements. 
To test hypothesis 3 a simple linear regression was performed, where the crowd-prediction is regressed on the random-walks prediction. If the crowd-prediction is not significantly different from the random-walks prediction, i.e. hypothesis 3 holds, than the two predictions should be perfectly correlated and the slope of the regression line should be 1 and the intercept 0. Equation \ref{OLS} describes the regression and $x$ corresponds to the prediction made by the random-walk. 

\begin{equation}\label{OLS}
    x = \beta_0 + \beta_1 \text{crowd} + \psi
\end{equation}

\section{Results}

\begin{figure}
    \centering
    \includegraphics[width=\textwidth]{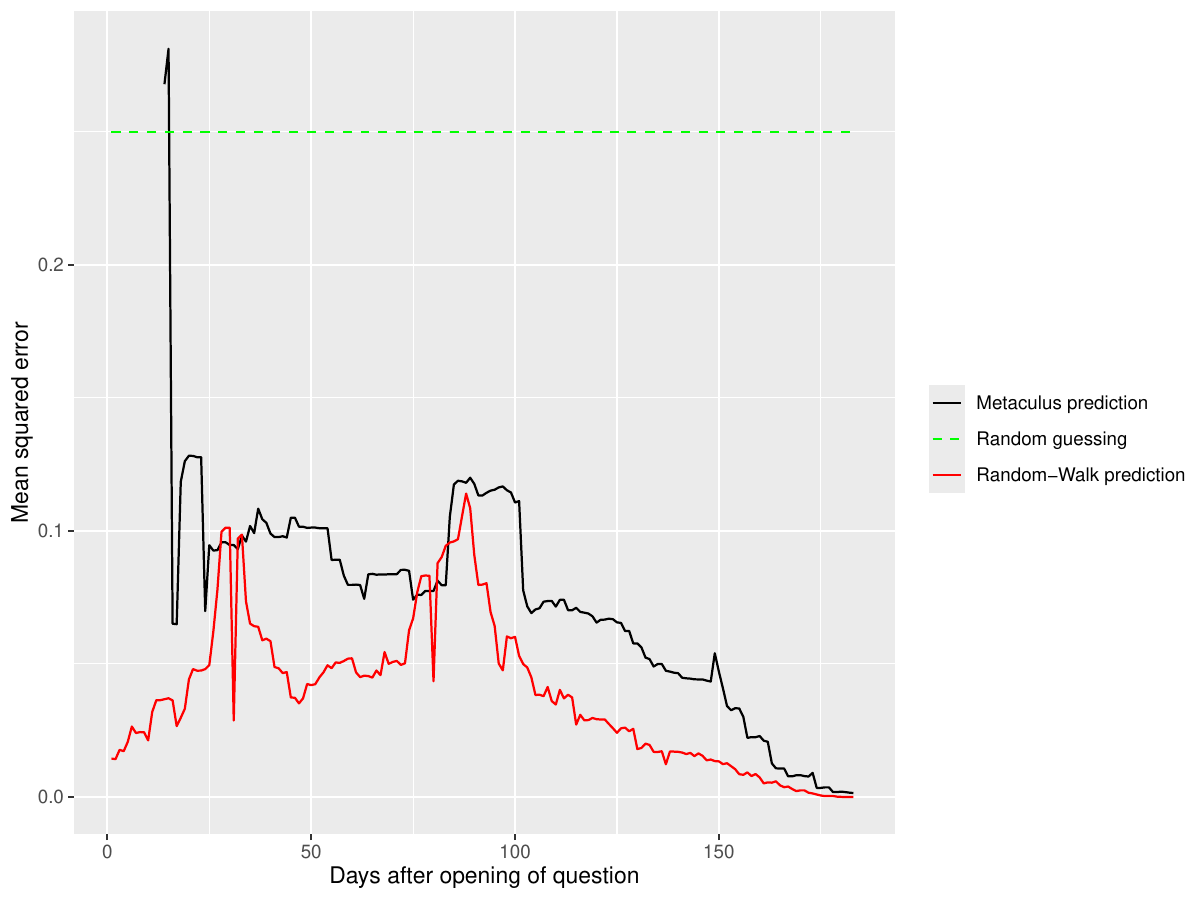}
    \caption{Accuracy of the Metaculus Prediction compared to the predictions made by the random-walk}
    \label{fig:mp_vs_rw_all}
\end{figure}

Figure \ref{fig:mp_vs_rw_all} plots the accuracy of the two methods over time, as measured by the mean squared error. The black line represents the evolving error of crowd-predictions over time (x-axis), while the red line shows the error of the random-walk model. The dashed green line represents the error that a random guesser would achieve. Initially, the error of crowd-predictions fluctuates substantially due to the limited number of predictions available at the start, resulting in a small sample size for the forecast combination method. As more predictions are received, the combined predictions and the error tend to change considerably.
As the figure \ref{fig:mp_vs_rw_all} already shows, the random-walk makes more accurate predictions on average.  Table \ref{DM_Test_hyp1} contains the full results of the test. The average error of the crowd-prediction is $0.0725$ and the average error of the random-walks prediction is $0.0421$. A relatively large difference in error of around $0.0304$ results.
The Diebold-Mariano test informs us that this difference is extremely unlikely to be a chance result.
Therefore, we need to reject hypothesis 1 and conclude that the random-walk provided far more accurate predictions.

\begin{table}[ht]
\centering
\caption{Diebold-Mariano test regarding hypothesis 1}
\label{DM_Test_hyp1}
\begin{tabular}{lcccc}
  \hline\hline
 & Estimate & Std. Error & $t$-value & $p$-value \\ 
  \hline
  avg. error difference & 0.0304168*** & 0.0048398 & 6.2847 & $<10^{-6}$\\
\hline
\end{tabular}
\end{table}

However, when compared with the strategy of random guessing, the crowd-prediction is significantly more accurate. 
The results regarding hypothesis 2 are printed in table \ref{DM_Test_hyp2}. 
The average error difference of $0.1775$ arises as the difference between the average error of the crowd and $0.25$. The hypothesis 2 is confirmed.

\begin{table}[ht]
\centering
\caption{Diebold-Mariano test regarding hypothesis 2}
\label{DM_Test_hyp2}
\begin{tabular}{lcccc}
  \hline\hline
 & Estimate & Std. Error & $t$-value & $p$-value \\ 
  \hline
  avg. error difference & 0.1775*** & 0.0078074 & 22.734 & $<10^{-6}$\\
\hline
\end{tabular}
\end{table}

Given that the crowd-prediction is less accurate than the random-walk, we may investigate hypothesis 3 and ask: Do both methods generate systematically different predictions? 
Table \ref{OLS regression} contains results of the regression analysis described in the previous section.
From them we conclude that the random-walks predictions are significantly different from the crowd-predictions. Specifically, the crowd-predictions are far higher on average.
The random-walk assigns on average only 84\% of the probability crowd-prediction to the currency depreciating, as the slope coefficient $crowd$ is 0.84.  
The intercept implies that the crowd-prediction is additionally unconditionally higher than the random-walks prediction, roughly by 1 percentage point.

\begin{table}[ht]
\centering
\caption{Regression results regarding hypothesis 3}
\label{OLS regression}
\begin{tabular}{lcccc}
  \hline\hline
 & Estimate & Std. Error & $t$-value & $p$-value \\ 
  \hline
  Intercept & -0.01199** & 0.00370 & -3.242 & 0.00121\\ 
  crowd & 0.84076*** & 0.01248 & -12.75962 & $<10^{-6}$\\ 
   \hline
   $R^2$ & 0.6998 &&             \\
\hline
\end{tabular}
\end{table}


%

\section{Discussion}

Using crowd-prediction for forecasting exchange rates should be avoided, if the random-walk is available,
because the latter yields more accurate predictions in this study. 
This study documents that the forecasting error of the crowd-prediction is a whopping 72\% higher, on average, when compared to the random-walk.

Crowd-predictions are increasingly incorporated as a decision-support, which has tremendous potential, but we should carefully assess whether better alternatives exist before doing so. 
In the realm of exchange rates, a statistical forecasting technique fares better. 
A lot of literature that touts the benefits of crowd-prediction, mostly along the line of research pursued by Tetlock (see e.g. \cite{tetlock2016superforecasting}). This result is not in contrast with this research, as the crowd did successfully forecast exchange rates, significantly outperforming random guessing. Yet, the random-walk is still better.

Should we expect the crowd and the random-walk to produce forecasts that can be combined to yield a better forecast? Probably not.
The crowd mostly seems to have produced forecasts that are mostly less certain and less responsive to exchange rate movements. Even on a question-by-question basis, the crowd is often clearly less accurate than the random-walk. 
The graphs in the appendix, which detail all predictions, provide great visual evidence of this. 


The studies results are to be interpreted in light of the setting. Since the forecasts were only 6-month ahead (at maximum), the prediction accuracy on long-term outcomes might be different.
Furthermore, since substantial currency depreciation is a rare phenomenon, this study involves some low probability forecasts, which are more difficult to interpret given the limited sample of events. However, this is not very relevant, as these predictions do not contribute much to the mean squared error and thus do not affect the outcome of this study much. 
We have no information regarding the forecasters knowledge of exchange rates, and thus it is not implausible to conjecture that a set of experts could have provided more accurate forecasts. 

\section{Conclusion}\label{sec13}

As crowd-prediction platforms increasingly inform public and private expectations regarding future events, their reliability and comparative efficacy warrants examination.
This paper compared the accuracy of exchange rate predictions from the Metaculus platform  to a well-studied statistical benchmark, the random-walk without drift, which provided mixed evidence. The crowd did possess considerable foresight but the random-walk without drift provided significantly more accurate predictions. 
Undoubtedly, the increasing use of crowd-predictions can improve decision-making in countless areas, and is thus a great development.
However, this analysis shows that simple statistical methods can be far more accurate than the crowd-prediction. Therefore, when it comes to decision-making, it is imperative to exercise caution in the use and interpretation of crowd-predictions and to evaluate whether more effective alternatives are available.
To better understand what level of confidence we should place into crowd-prediction, and when to utilize crowd-prediction over other methods, we need more research. We should e.g. analyze which factors predict accuracy, so that we can know when to expect reliable forecasts.
While this study specifically examines exchange rate forecasts, extending similar analyses to other domains would reveal how crowd-prediction fares relative to other forecasting techniques. 
Currently, this line of research is inhibited by a limited sample of available crowd-predictions. Thus, another avenue for future research is to collect crowd-predictions on events that allow a rigorous comparison to other means of collecting forecasts.

\newpage
\printbibliography


\setcounter{figure}{0}
\renewcommand{\thefigure}{A\arabic{figure}}

\newpage
\section*{Appendix}

\begin{figure}[h]
    \centering
    \includegraphics[width=.8\textwidth]{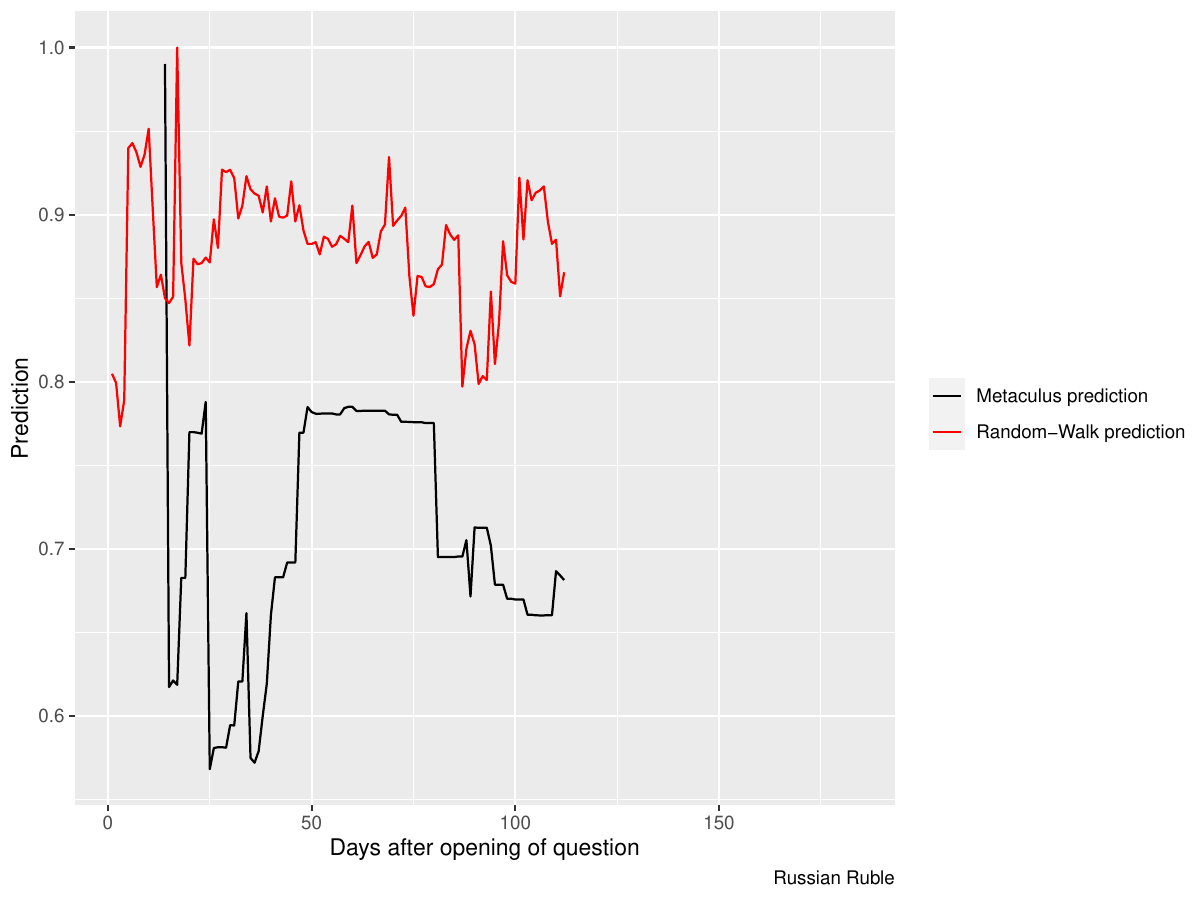}
    \includegraphics[width=.8\textwidth]{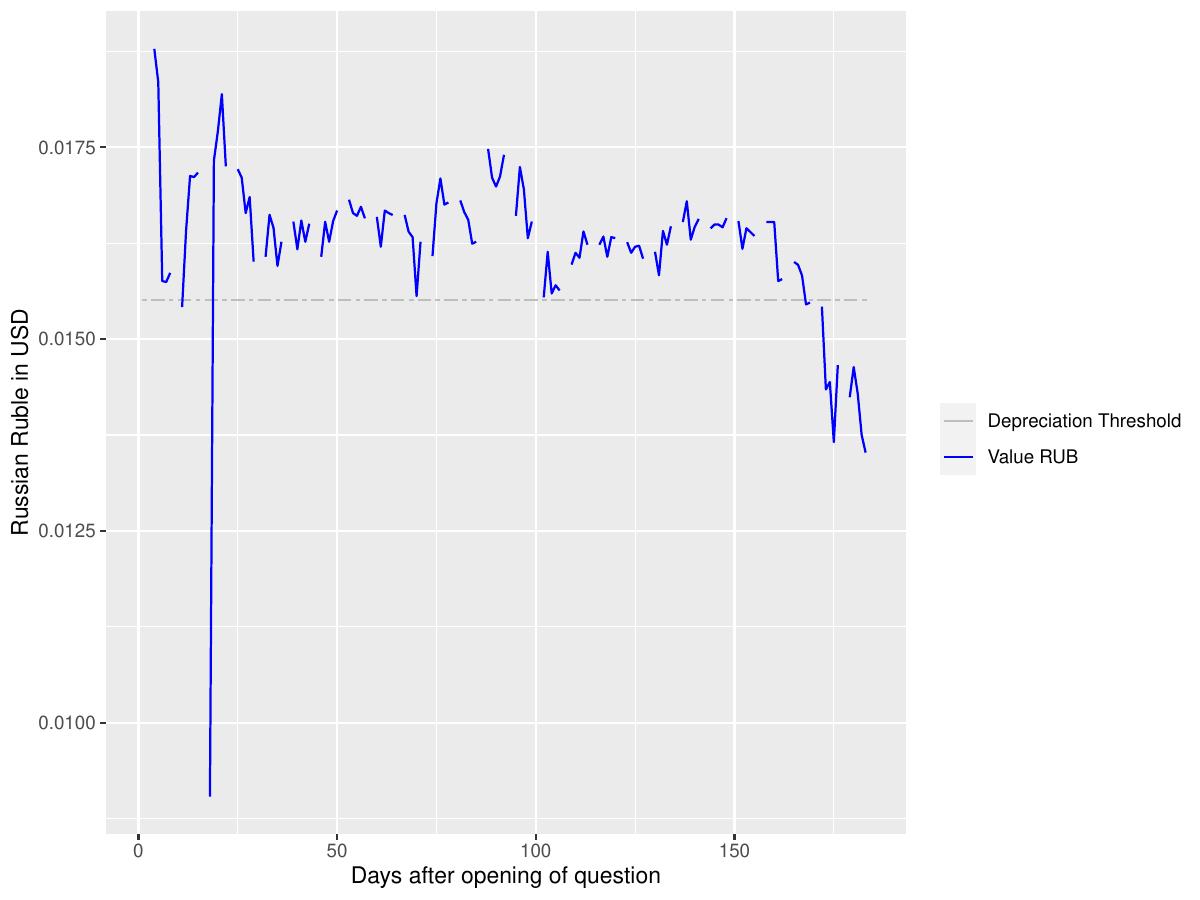}
    \caption{Top picture: Metaculus Prediction compared to the predictions made by the random-walk \\ 
    Bottom picture: Exchange rate across the same time} 
    \label{fig:mp_vs_rw_RUB}
\end{figure}

According to figure \ref{fig:mp_vs_rw_RUB} the Russian Ruble triggers the depreciation threshold very early. There are two important remarks to be made here: First, the data from the quantmod-package records a downward spike at around day 20 that is not present in the data from xe.com and therefore would not have triggered the question to resolve positively at that time. However, the data from xe.com records the Russian Ruble crossing the depreciation threshold around day 100, whereas the Ruble barely touches the threshold in the quantmod-data. Secondly, both databases record the Ruble going below the depreciation threshold on day 13. However, predictions were just collected from day 14 onwards. Therefore, the question did not resolve as `Yes` $k=1$ on day 13, as the tournament had not really started at that point.

\begin{figure}[h]
    \centering
    \includegraphics[width=.8\textwidth]{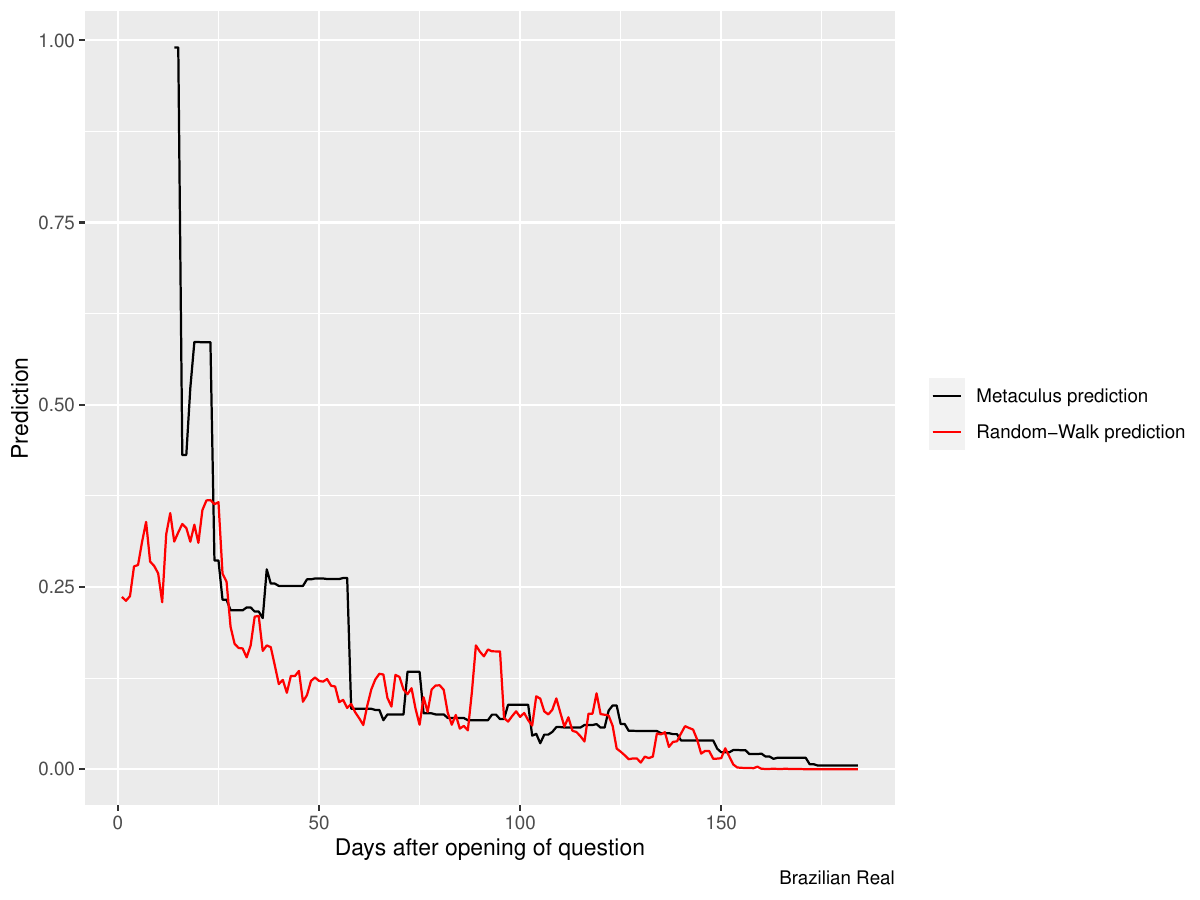}
    \includegraphics[width=.8\textwidth]{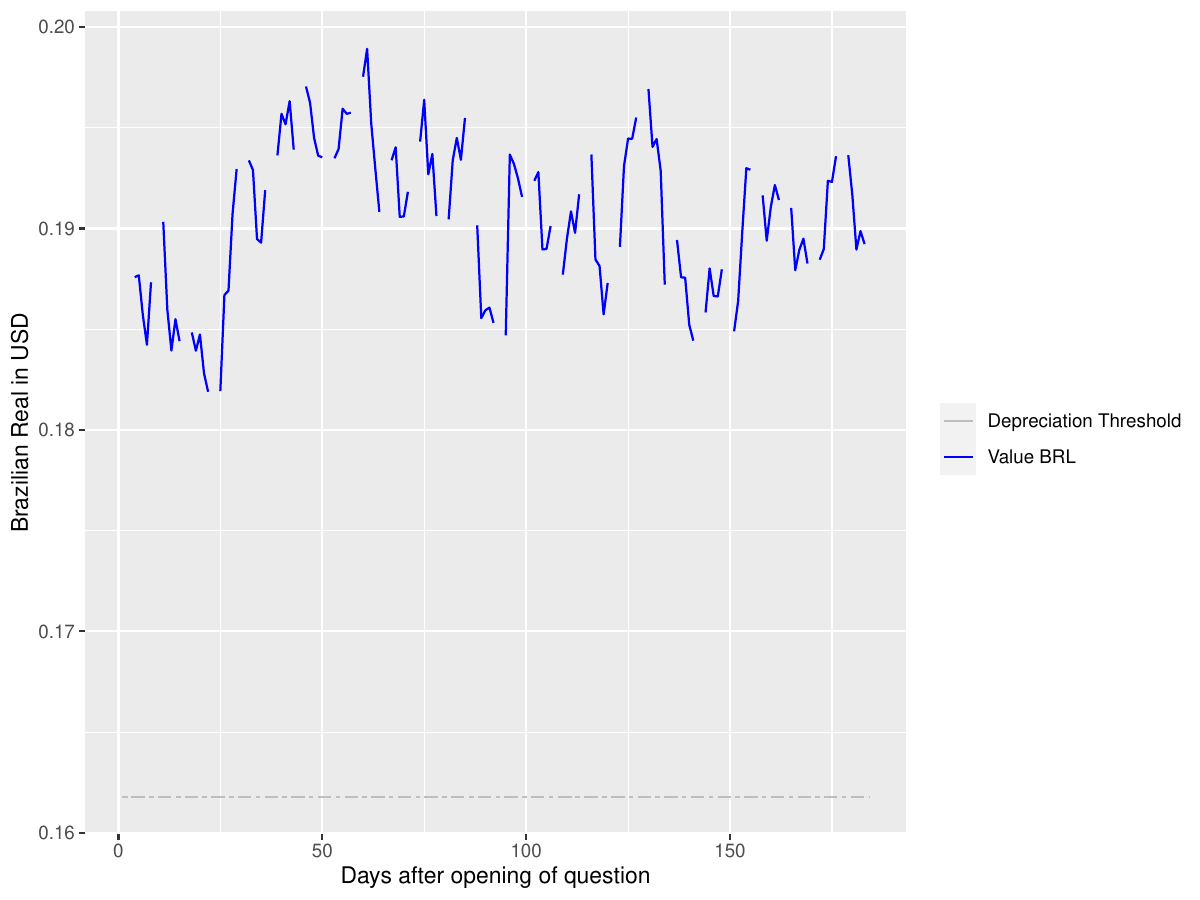}
    \caption{Top picture: Metaculus Prediction compared to the predictions made by the random-walk \\ 
    Bottom picture: Exchange rate across the same time}
    \label{fig:mp_vs_rw_BRL}
\end{figure}

\begin{figure}[h]
    \centering
    \includegraphics[width=.8\textwidth]{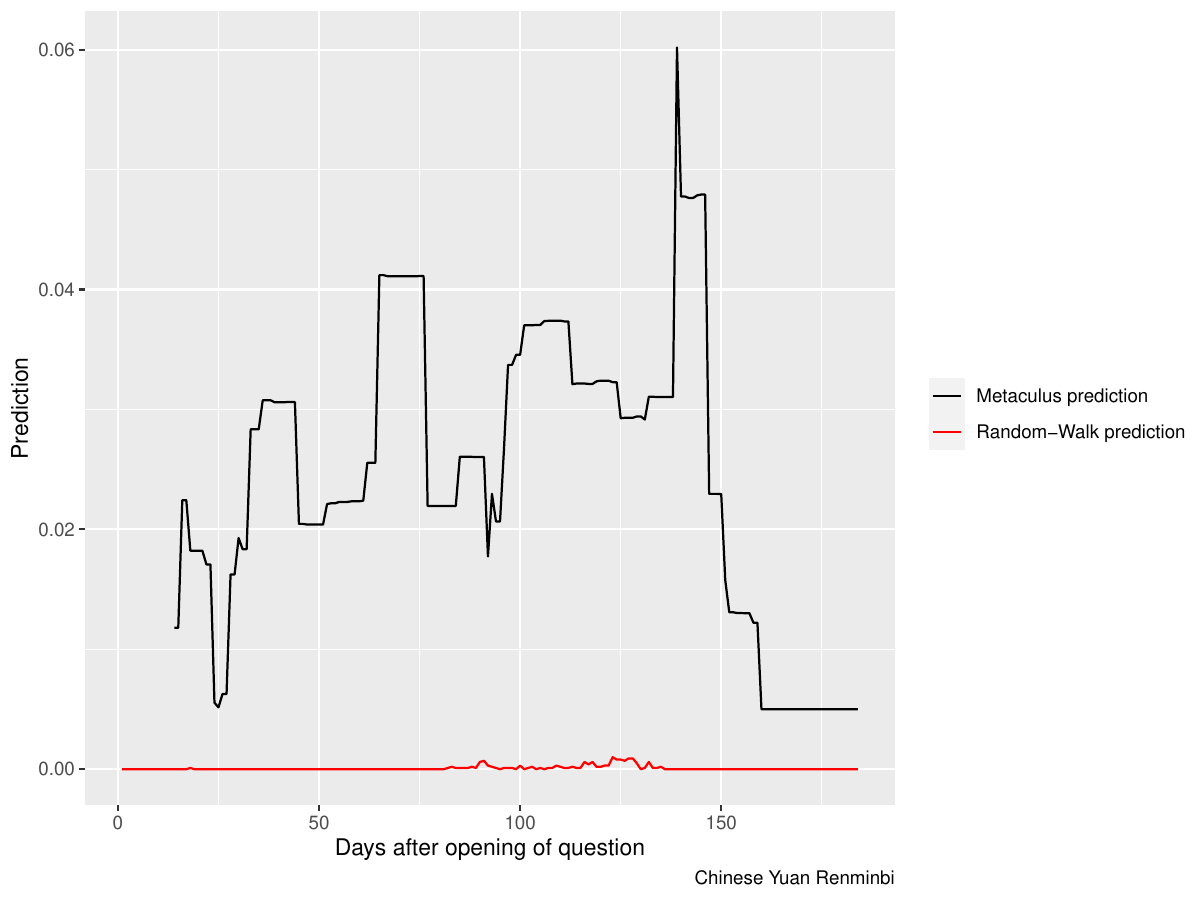}
    \includegraphics[width=.8\textwidth]{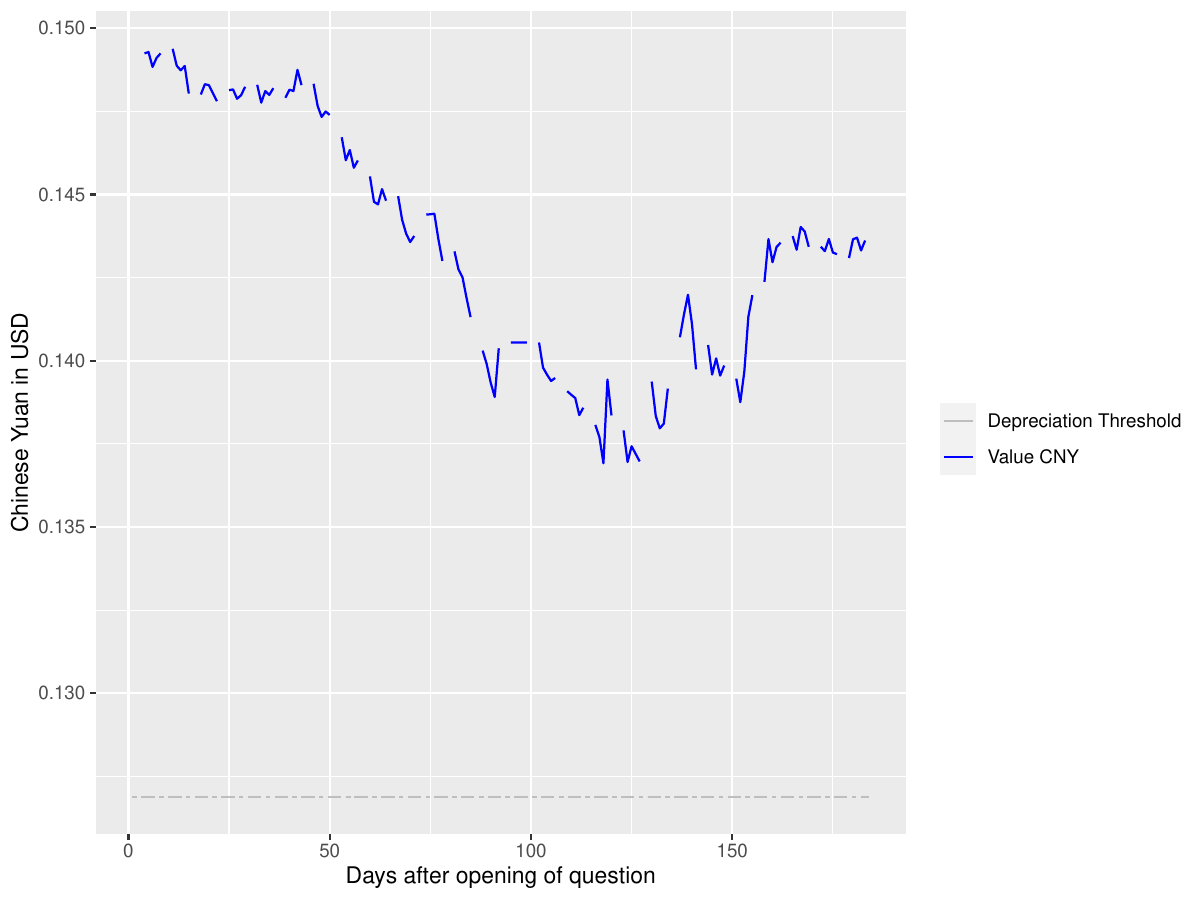}
    \caption{Top picture: Metaculus Prediction compared to the predictions made by the random-walk \\ 
    Bottom picture: Exchange rate across the same time}
    \label{fig:mp_vs_rw_CNY}
\end{figure}

\begin{figure}[h]
    \centering
    \includegraphics[width=.8\textwidth]{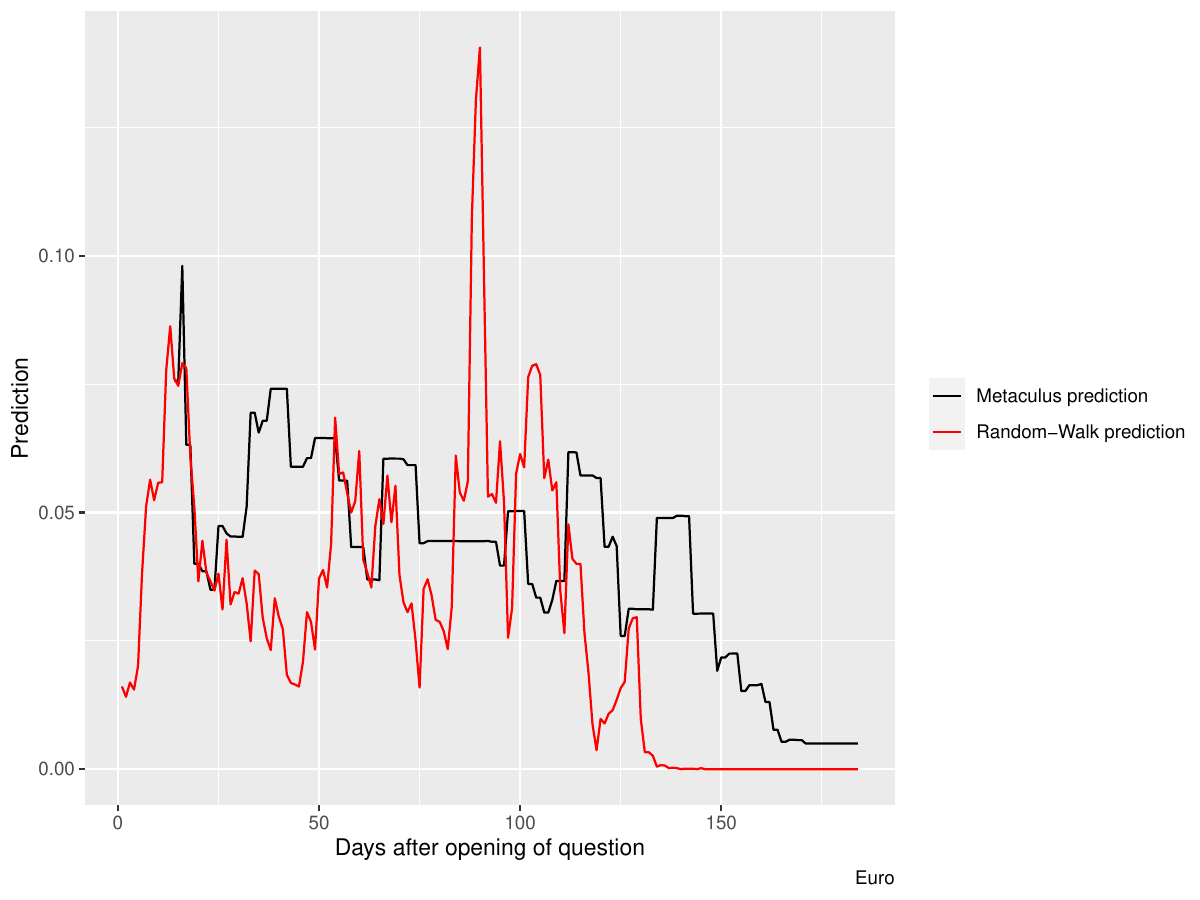}
    \includegraphics[width=.8\textwidth]{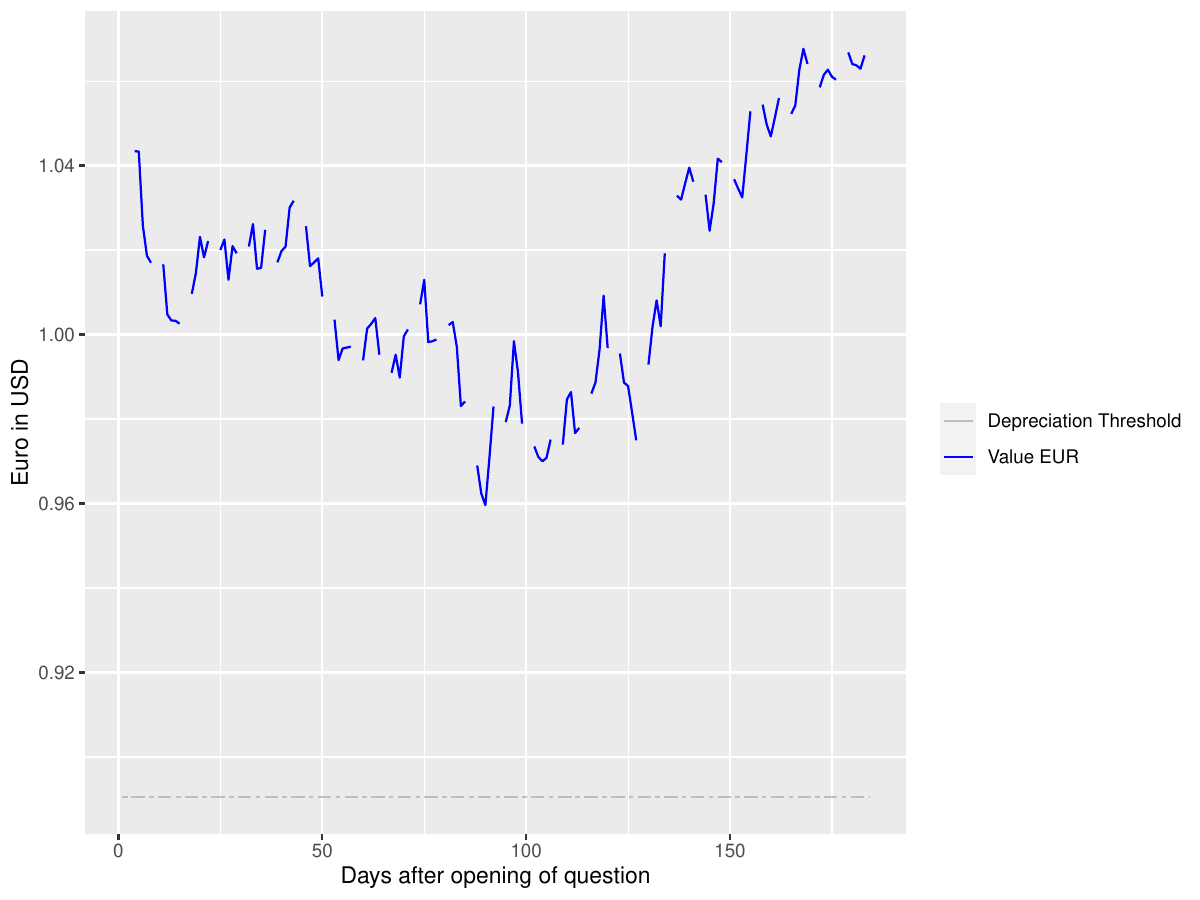}
    \caption{Top picture: Metaculus Prediction compared to the predictions made by the random-walk \\ 
    Bottom picture: Exchange rate across the same time}
    \label{fig:mp_vs_rw_EUR}
\end{figure}

\begin{figure}[h]
    \centering
    \includegraphics[width=.8\textwidth]{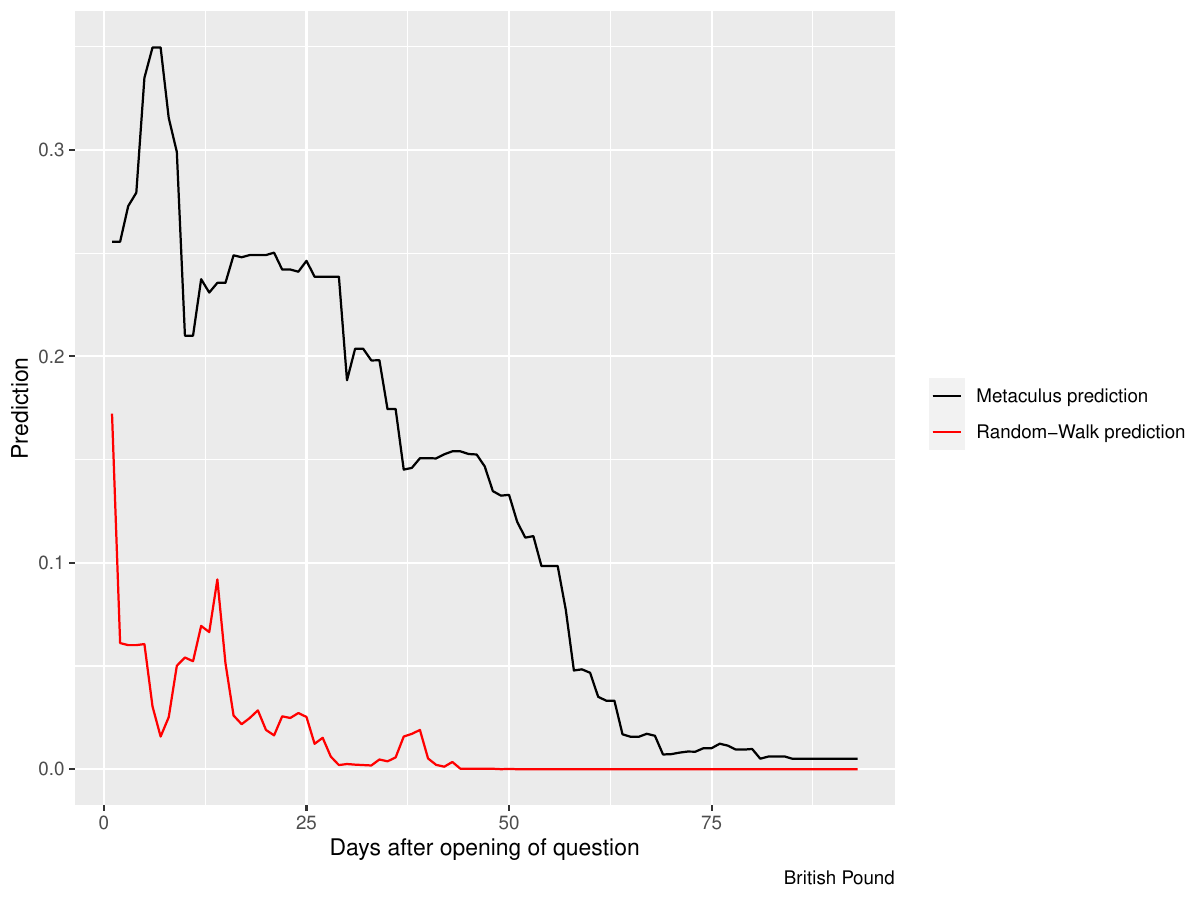}
    \includegraphics[width=.8\textwidth]{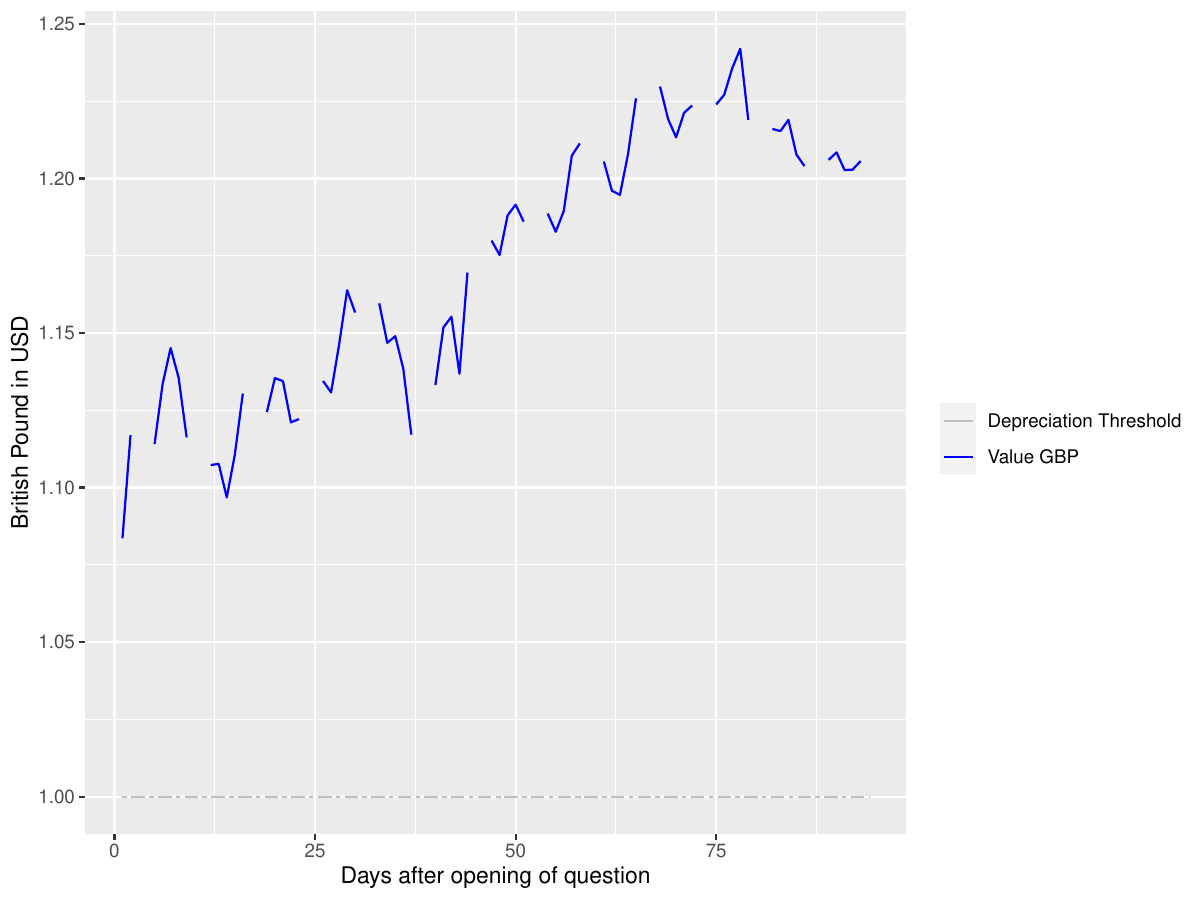}
    \caption{Top picture: Metaculus Prediction compared to the predictions made by the random-walk \\ 
    Bottom picture: Exchange rate across the same time}
    \label{fig:mp_vs_rw_GBP}
\end{figure}

\begin{figure}[h]
    \centering
    \includegraphics[width=.8\textwidth]{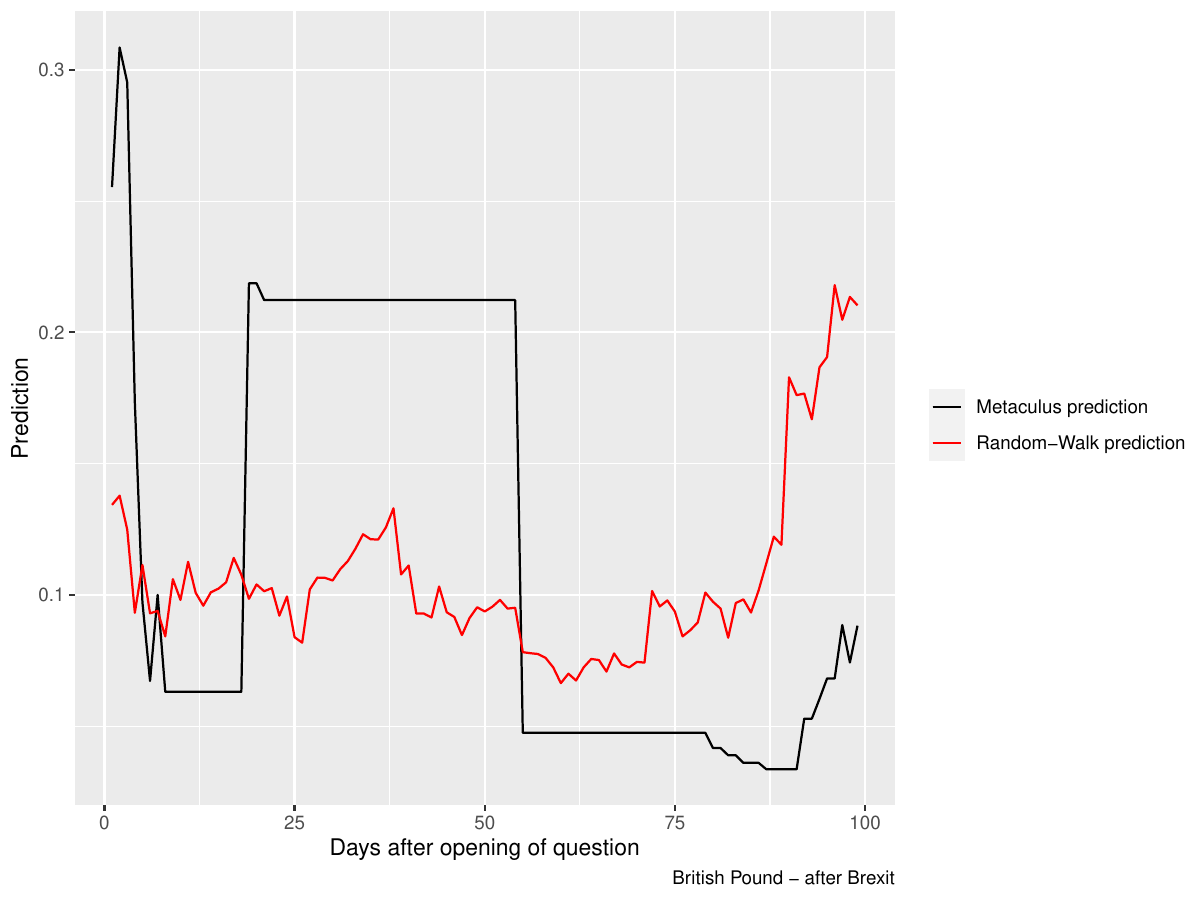}
    \includegraphics[width=.8\textwidth]{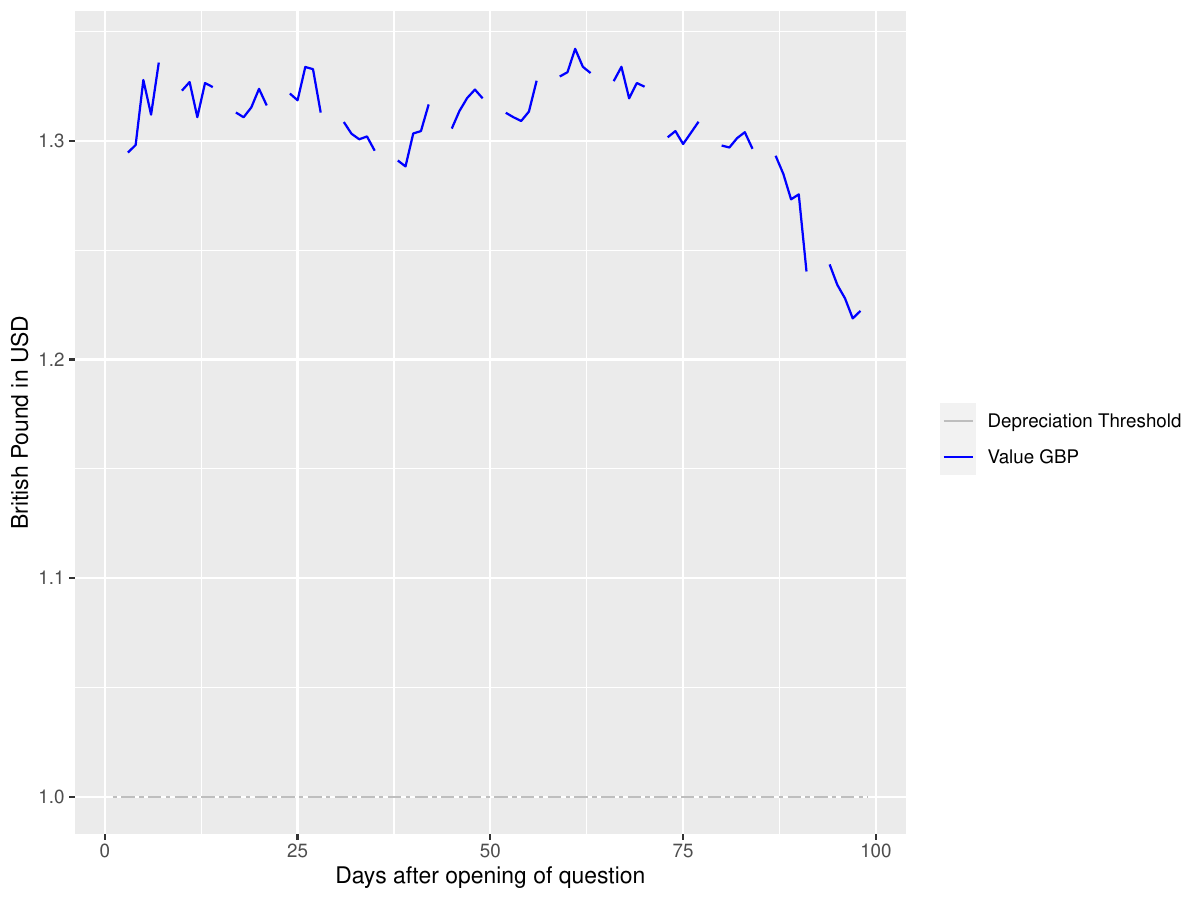}
    \caption{Top picture: Metaculus Prediction compared to the predictions made by the random-walk \\ 
    Bottom picture: Exchange rate across the same time}
    \label{fig:mp_vs_rw_GBP_2}
\end{figure}

\begin{figure}[h]
    \centering
    \includegraphics[width=.8\textwidth]{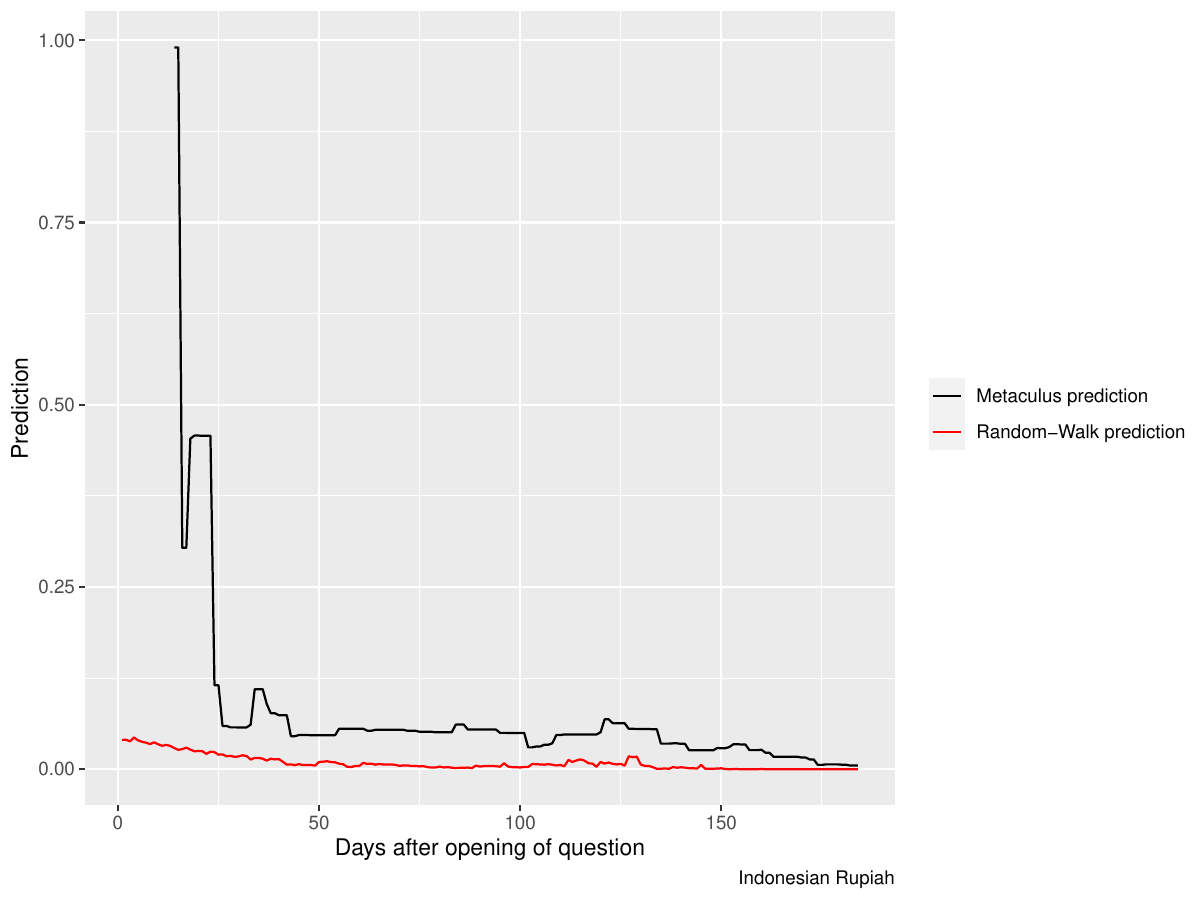}
    \includegraphics[width=.8\textwidth]{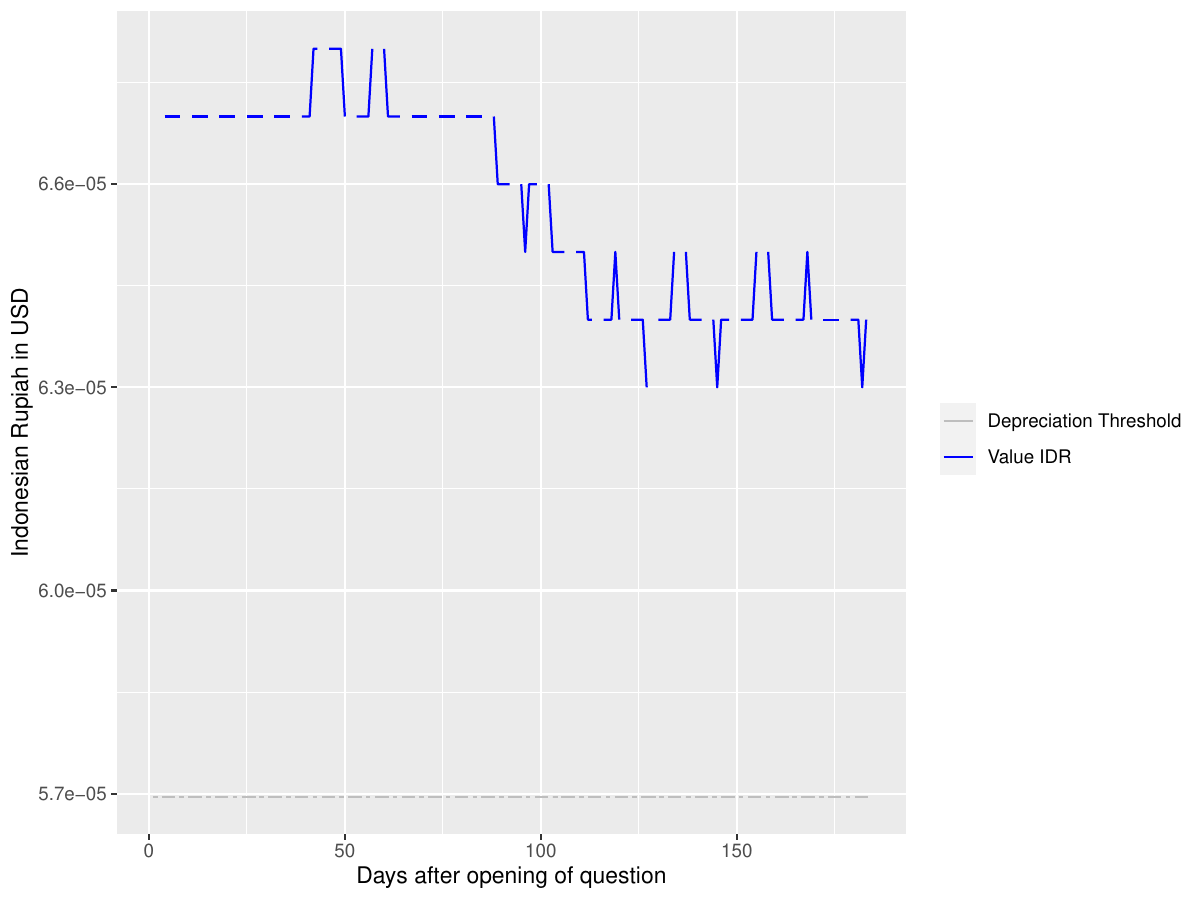}
    \caption{Top picture: Metaculus Prediction compared to the predictions made by the random-walk \\ 
    Bottom picture: Exchange rate across the same time}
    \label{fig:mp_vs_rw_IDR}
\end{figure}

\begin{figure}[h]
    \centering
    \includegraphics[width=.8\textwidth]{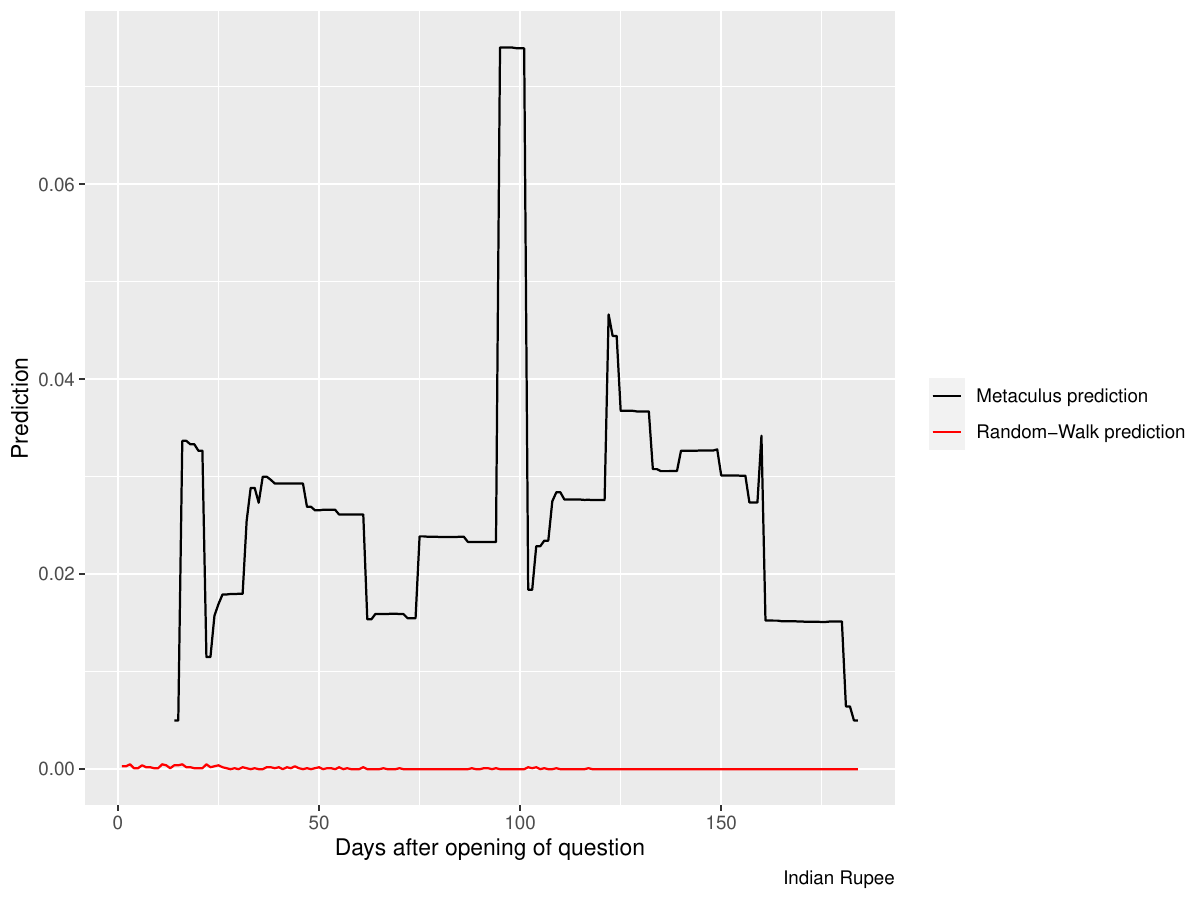}
    \includegraphics[width=.8\textwidth]{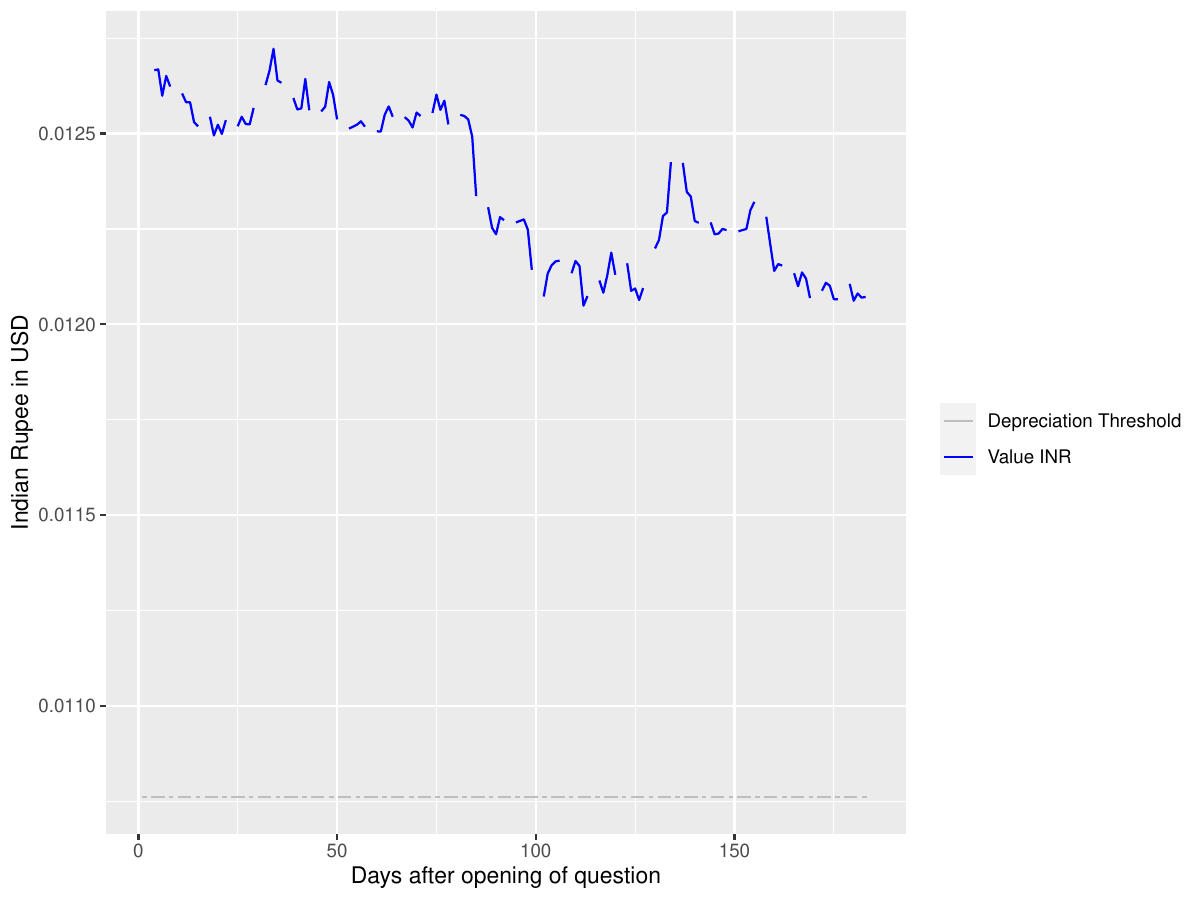}
    \caption{Top picture: Metaculus Prediction compared to the predictions made by the random-walk \\ 
    Bottom picture: Exchange rate across the same time}
    \label{fig:mp_vs_rw_INR}
\end{figure}

\begin{figure}[h]
    \centering
    \includegraphics[width=.8\textwidth]{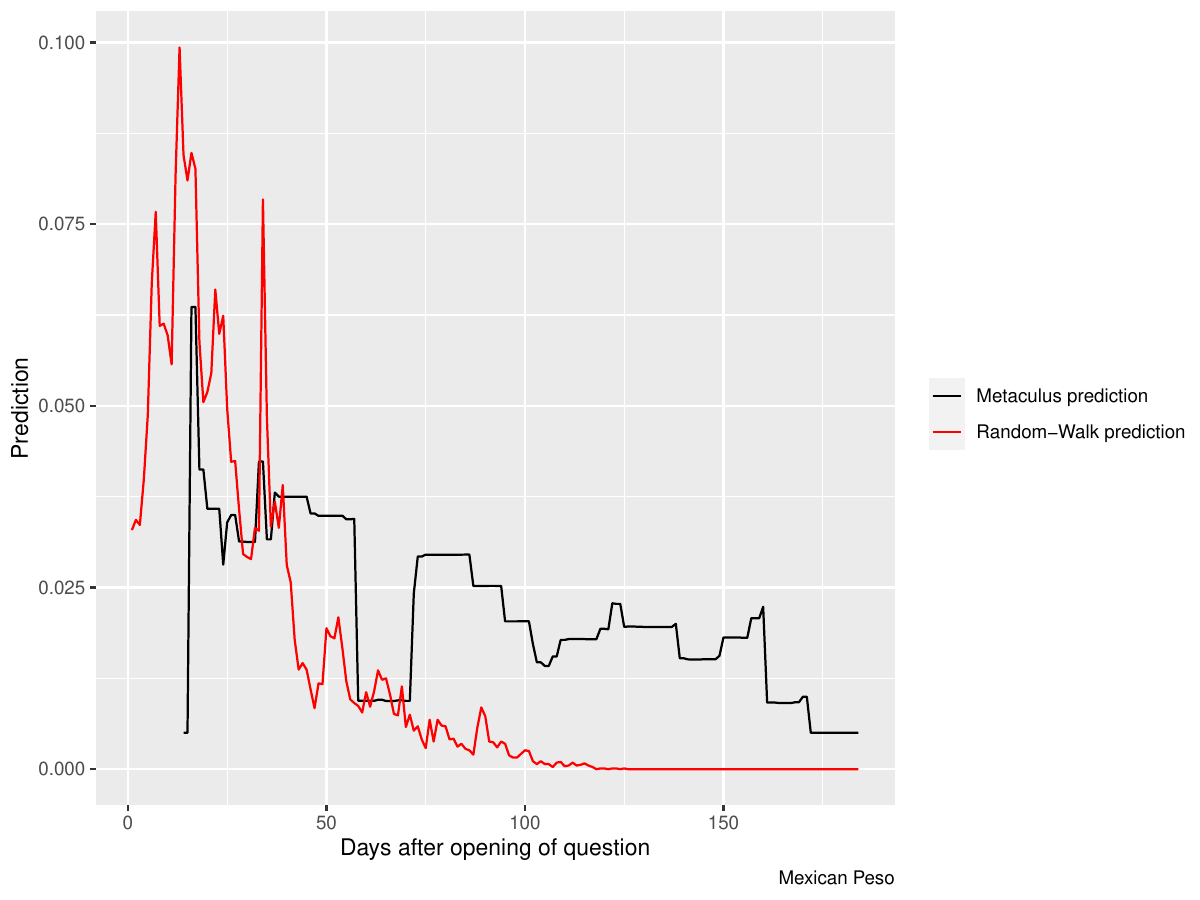}
    \includegraphics[width=.8\textwidth]{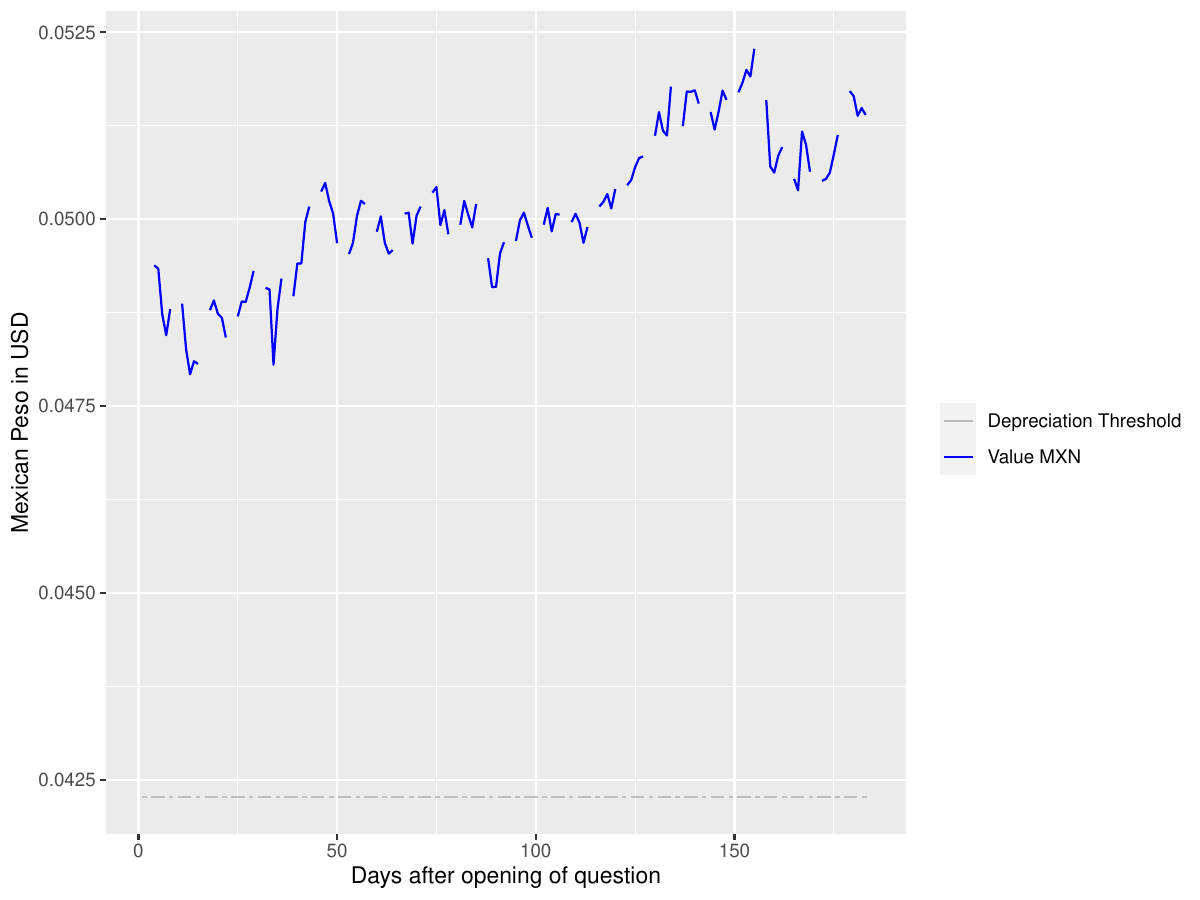}
    \caption{Top picture: Metaculus Prediction compared to the predictions made by the random-walk \\ 
    Bottom picture: Exchange rate across the same time}
    \label{fig:mp_vs_rw_MXN}
\end{figure}

\begin{figure}[h]
    \centering
    \includegraphics[width=.8\textwidth]{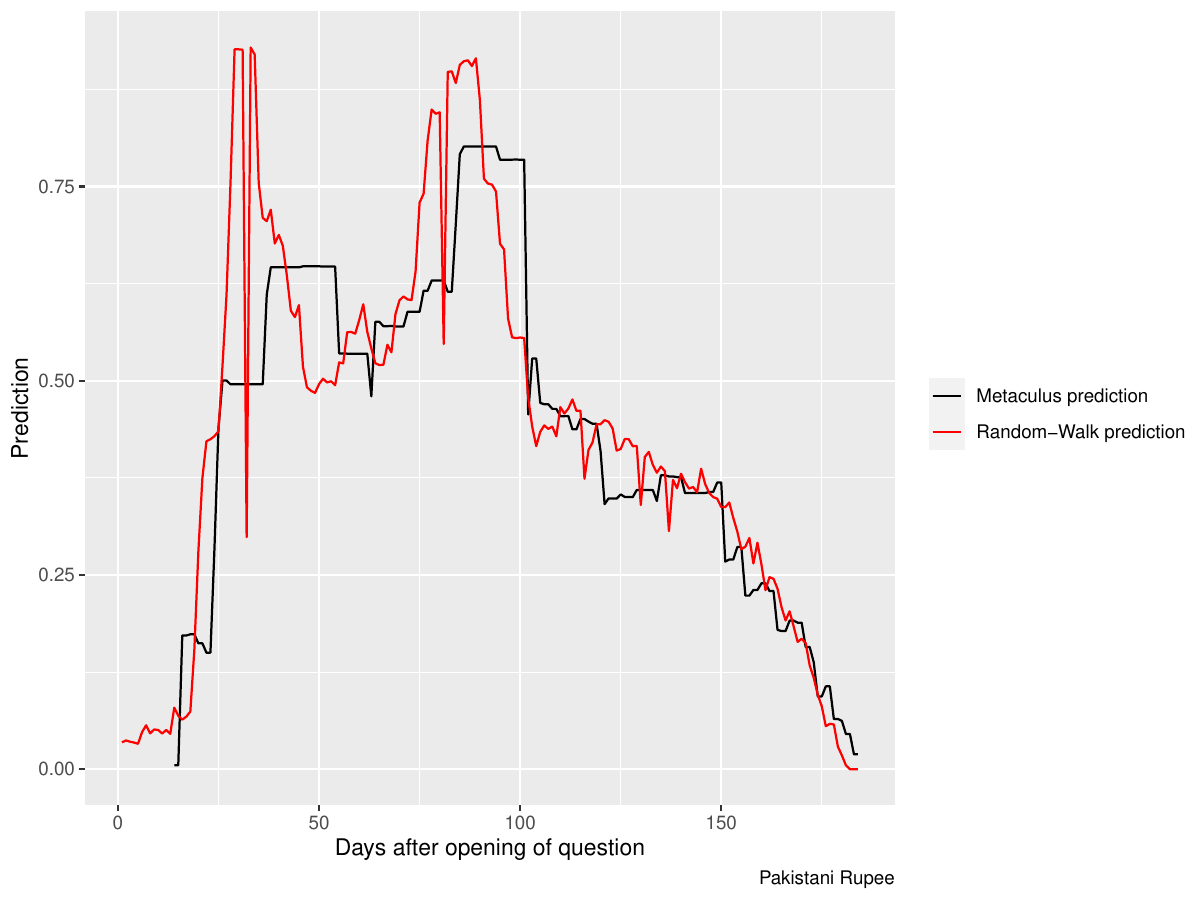}
    \includegraphics[width=.8\textwidth]{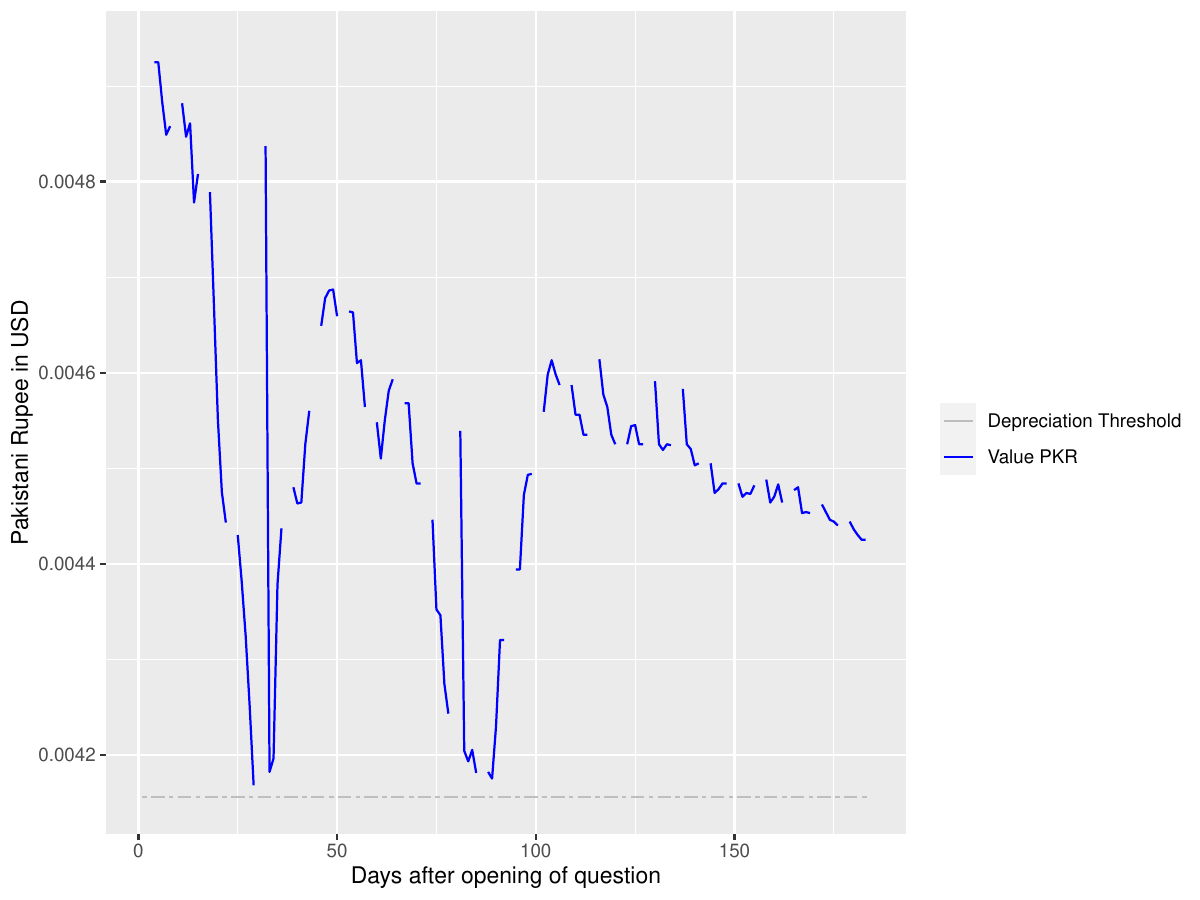}
    \caption{Top picture: Metaculus Prediction compared to the predictions made by the random-walk \\ 
    Bottom picture: Exchange rate across the same time}
    \label{fig:mp_vs_rw_PKR}
\end{figure}

\begin{figure}[h]
    \centering
    \includegraphics[width=.8\textwidth]{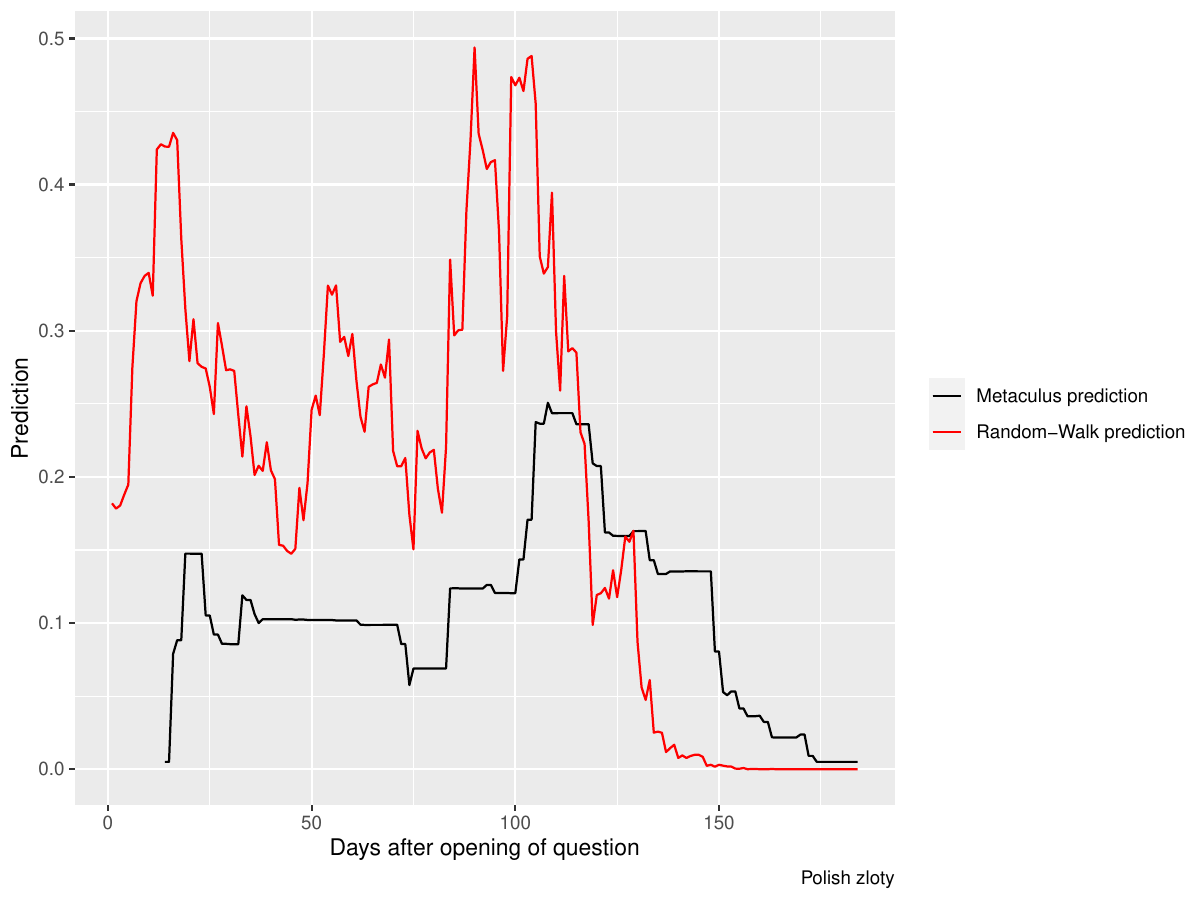}
    \includegraphics[width=.8\textwidth]{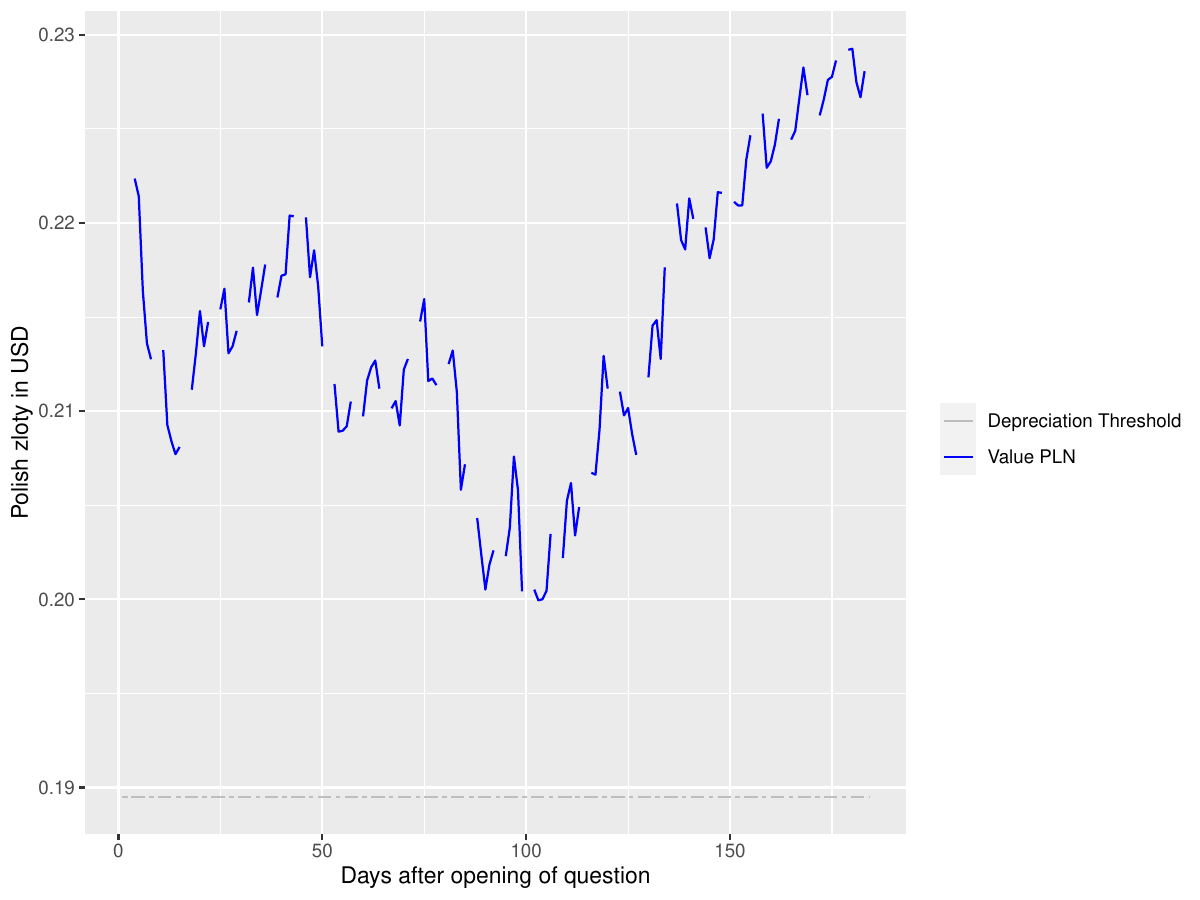}
    \caption{Top picture: Metaculus Prediction compared to the predictions made by the random-walk \\ 
    Bottom picture: Exchange rate across the same time}
    \label{fig:mp_vs_rw_PLN}
\end{figure}

\begin{figure}[h]
    \centering
    \includegraphics[width=.8\textwidth]{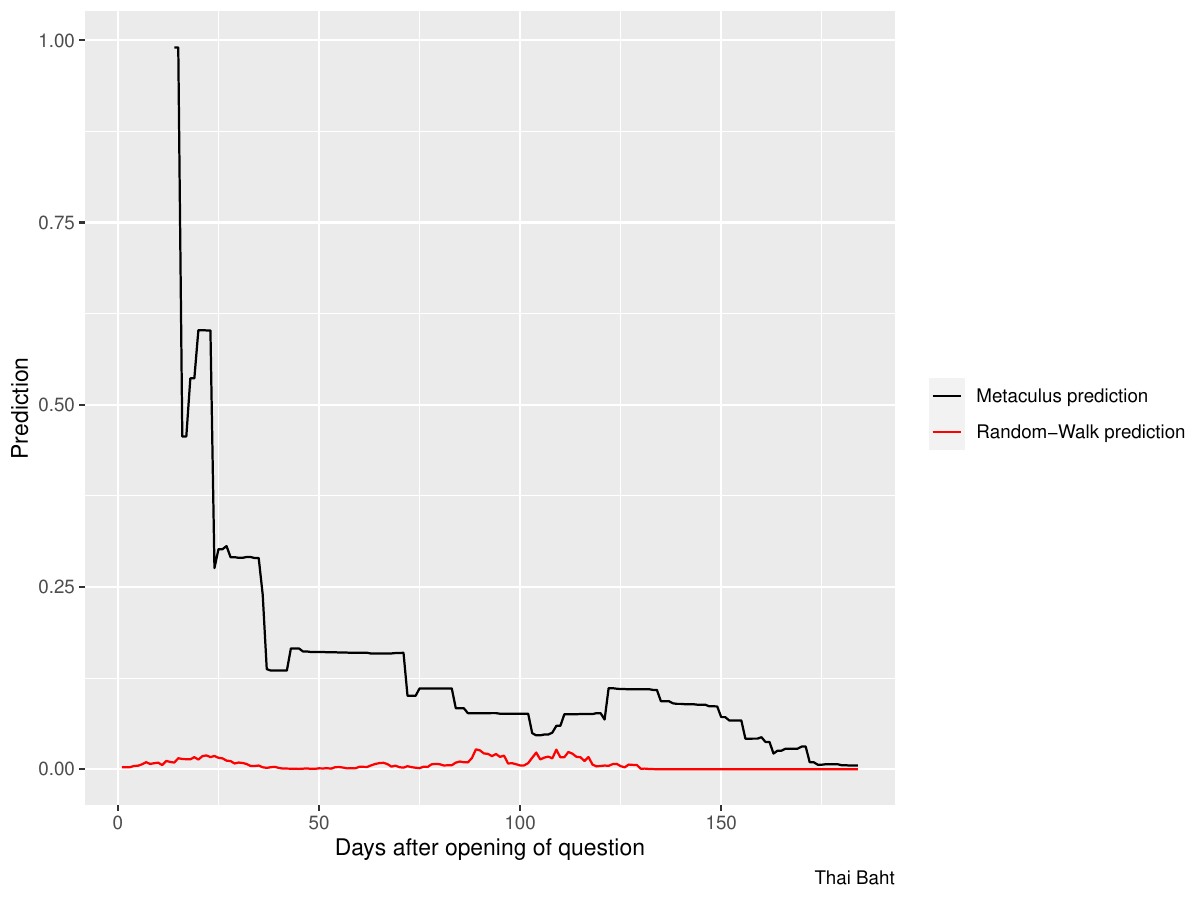}
    \includegraphics[width=.8\textwidth]{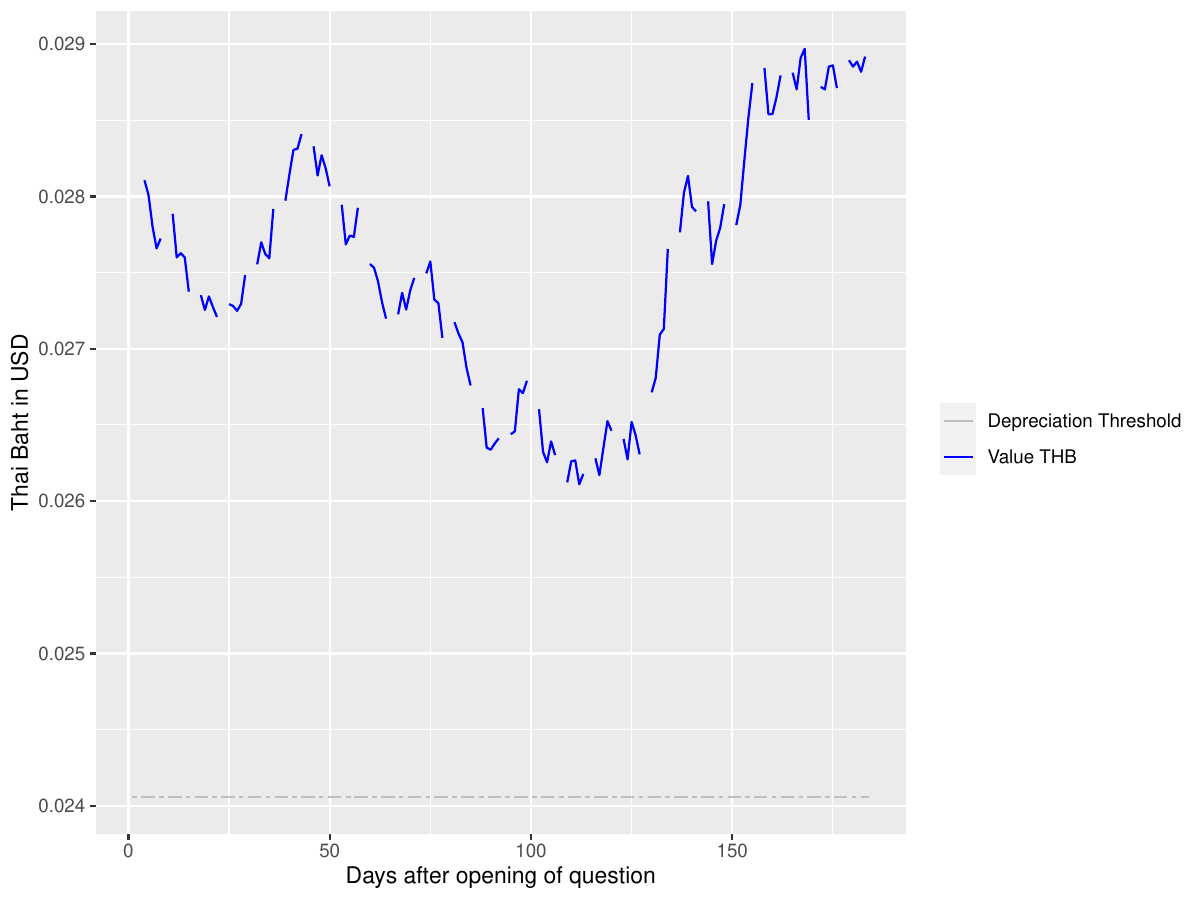}
    \caption{Top picture: Metaculus Prediction compared to the predictions made by the random-walk \\ 
    Bottom picture: Exchange rate across the same time}
    \label{fig:mp_vs_rw_THB}
\end{figure}

\begin{figure}[h]
    \centering
    \includegraphics[width=.8\textwidth]{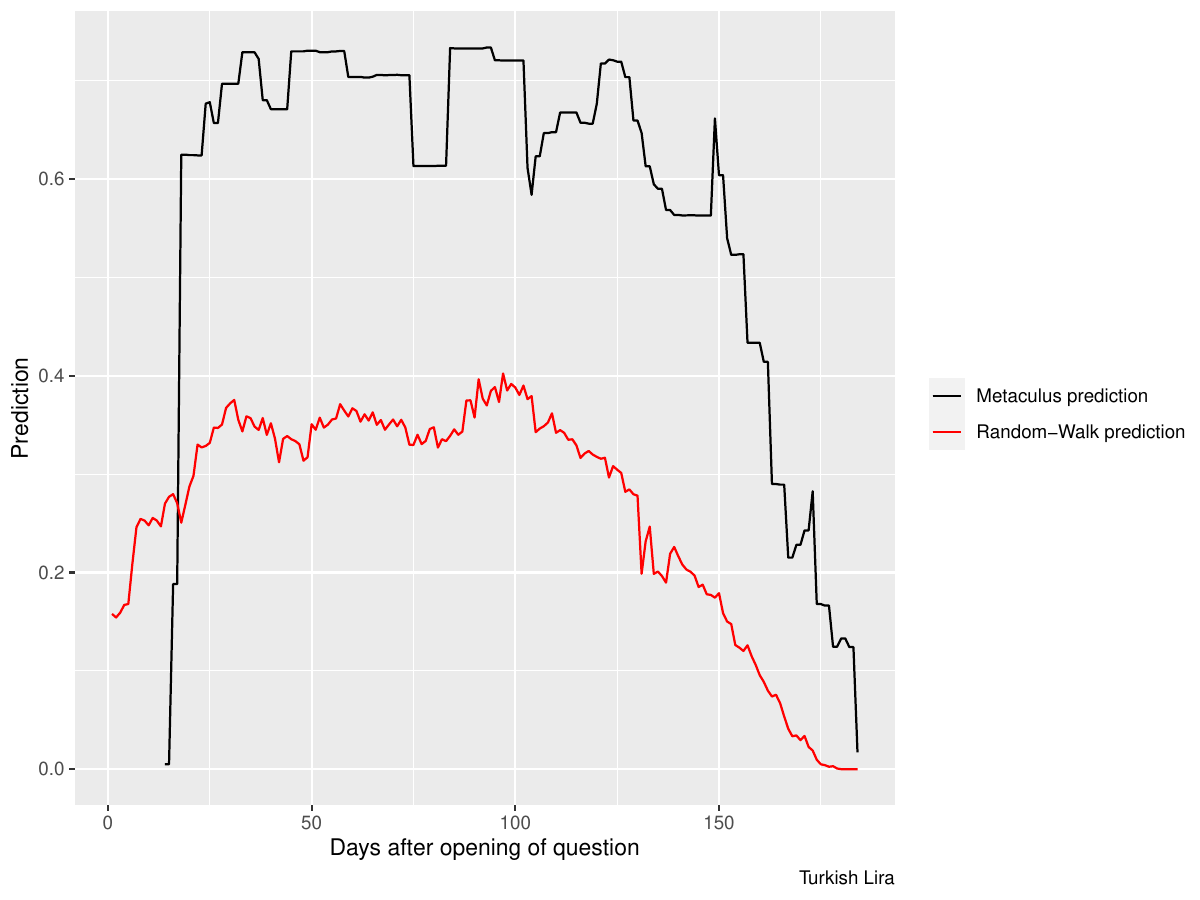}
    \includegraphics[width=.8\textwidth]{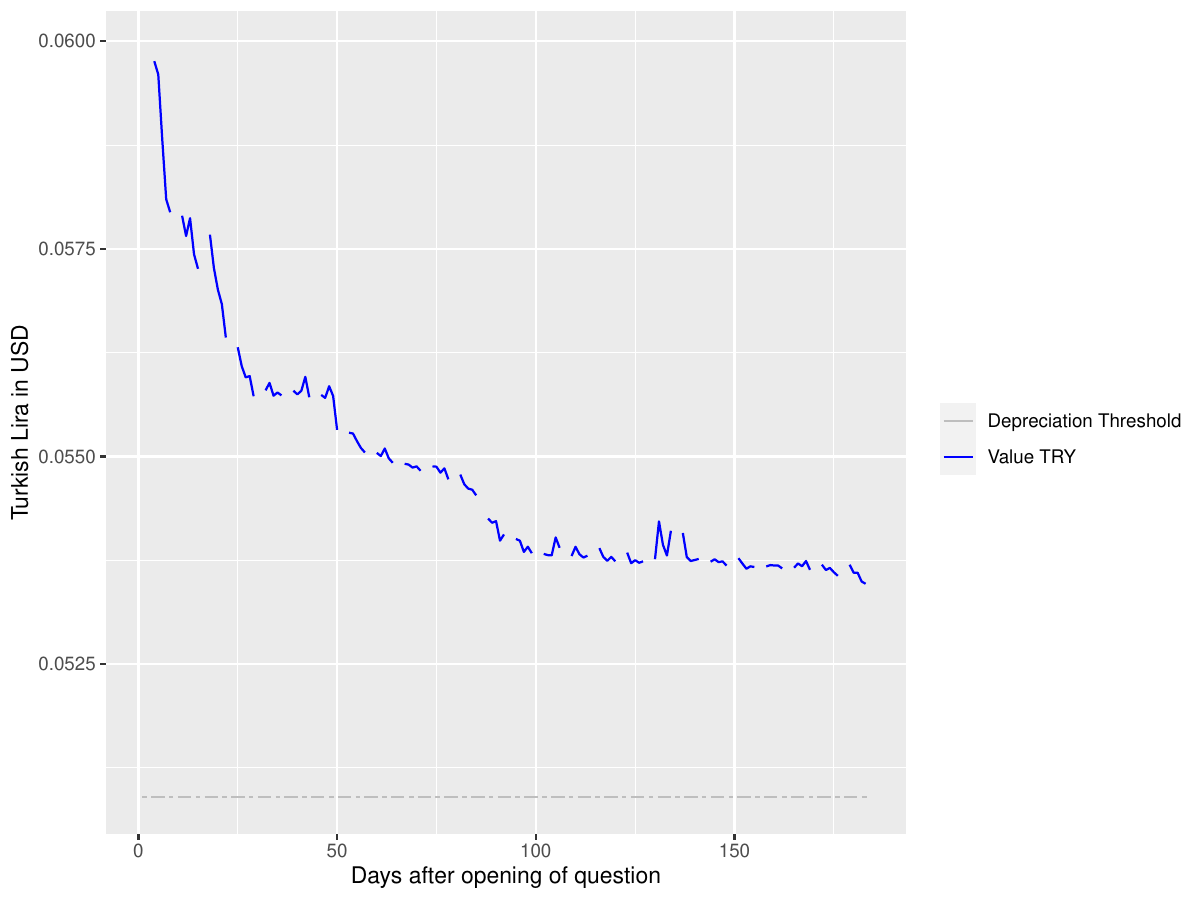}
    \caption{Top picture: Metaculus Prediction compared to the predictions made by the random-walk \\ 
    Bottom picture: Exchange rate across the same time}
    \label{fig:mp_vs_rw_TRY}
\end{figure}

\end{document}